\documentstyle[epsf]{article}
\newskip\humongous \humongous=0pt plus 1000pt minus 1000pt

\newif\ifdtup

\catcode`\@=11
\@addtoreset{equation}{section}
\def\theequation{\thesection\arabic{equation}}

\def\@normalsize{\@setsize\normalsize{15pt}\xiipt\@xiipt
\abovedisplayskip 14pt plus3pt minus3pt%
\belowdisplayskip \abovedisplayskip
\abovedisplayshortskip \z@ plus3pt%
\belowdisplayshortskip 7pt plus3.5pt minus0pt}

\def\small{\@setsize\small{13.6pt}\xipt\@xipt
\abovedisplayskip 13pt plus3pt minus3pt%
\belowdisplayskip \abovedisplayskip
\abovedisplayshortskip \z@ plus3pt%
\belowdisplayshortskip 7pt plus3.5pt minus0pt
\def\@listi{\parsep 4.5pt plus 2pt minus 1pt
     \itemsep \parsep
     \topsep 9pt plus 3pt minus 3pt}}

\relax
\catcode`@=12
\catcode`\@=11

\def\section{\@startsection{section}{1}{\z@}{3.5ex plus 1ex minus
   .2ex}{2.3ex plus .2ex}{\large\bf}}

\def\thesection{\arabic{section}.}

\def\appendix{\setcounter{section}{0}
 \def\thesection{Appendix \Alph{section}:}
 \def\theequation{\Alph{section}.\arabic{equation}}}

\begin{document}
\begin{titlepage}
\begin{center}
{\Large  
CP, Charge Fractionalizations and  Low Energy Effective Actions\\
in the $SU(2)$ Seiberg-Witten Theories with  Quarks
}
\end{center}
\vspace{1em}
\begin{center}
{\large Kenichi Konishi$^{(1)}$ and Haruhiko Terao$^{(2) \dagger}$ }
\end{center}
\vspace{1em}
\begin{center}
{\it 
Dipartimento di Fisica -- Universit\`a di Genova$^{(1)}$\\
Istituto Nazionale di Fisica Nucleare -- Sezione di Genova$^{(1,2)}$\\
Consiglio Nazionale delle Ricerche --  Area di Ricerca  di Genova$^{(2)}$ \\
Via Dodecaneso, 33 -- 16146 Genova (Italy)\\
E-mail: konishi@infn.ge.infn.it; terao@dipfis.ge.infn.it 
}
\end{center}
\vspace{3em}
\noindent
{\bf ABSTRACT:}
Several dynamical aspects of the $SU(2)$ Seiberg-Witten models with
$N_f$ quark hypermultiplets are explored. We first clarify the 
meaning of the number of the singularities of the space of vacua. 
CP properties  of the theories are then studied and periodicity of 
theories in $\theta$  with and without bare quark masses is  obtained
($(4- N_f) \pi$ and $\pi$, respectively). CP noninvariance at a generic
point of QMS manifests itself as the electric and quark-number charge
fractionalizations for the dyons; we show that the exact Seiberg-Witten 
solution contains such effects correctly, in agreement with the 
semiclassical analysis  recently made by F.Ferrari. Upon $N=1$ 
perturbation the low energy effective theories at the singularities 
display confinement, and in most cases chiral symmetry breaking as a 
consequence. In one of the vacua for $N_f=3$ confinement is not 
accompanied by chiral symmetry breaking: we interpret it as an
example of oblique confinement of 't Hooft. We discuss further the
consistency of the physical picture found here by studying the effects 
of soft supersymmetry breaking as well as the behavior of the theory in 
the $N=1$ limit.
\vspace{1.5em}
\begin{flushleft}
GEF-Th-6/1997; 
KANAZAWA 97-11\\
~~~~~~~~~~~~~~~~~~~~~~~~~~~~~~~~~~~~~~~~~~~~~~~~~~~~~~~~
~~~~~~~~~~~~~~~~~~~~~~~
~~~~~~~~~~~~~~~July 1997 
\end{flushleft}
\begin{flushleft}
$^{\dagger}$ 
On leave from and the address from 1/8/1997:
Department of Physics, \\ {\hskip 2 mm} Kanazawa University, Kanazawa, Japan.
\end{flushleft}
\end{titlepage}
\newcommand{\beq}{\begin{equation}}
\newcommand{\eeq}{\end{equation}}
\newcommand{\bea}{\begin{eqnarray}}
\newcommand{\eea}{\end{eqnarray}}
\newcommand{\beas}{\begin{eqnarray*}}
\newcommand{\eeas}{\end{eqnarray*}}
\newcommand{\defi}{\stackrel{\rm def}{=}}
\newcommand{\non}{\nonumber}
\def\dirac{{\cal D}}
\def\dplus{{\cal D_{+}}}
\def\dminus{{\cal D_{-}}}
\def\dbar{\bar{D}}
\def\H{\cal{H}}
\def\de{\partial}
\def\si{\sigma}
\def\sb{{\bar \sigma}}
\def\rn{{\bf R}^n}
\def\r4{{\bf R}^4}
\def\s4{{\bf S}^4}
\def\ker{\hbox{\rm ker}}
\def\dim{\hbox{\rm dim}}
\def\sup{\hbox{\rm sup}}
\def\inf{\hbox{\rm inf}}
\def\infi{\infty}
\def\nrm{\parallel}
\def\nrmi{\parallel_\infty}
\def\om{\Omega}
\def\Tr{ \hbox{\rm Tr}}
\def\const{\hbox {\rm const.}}
\def\o{\over}
\def\th{\theta}
\def\im{\hbox{\rm Im}}
\def\re{\hbox{\rm Re}}
\def\bra{\langle}
\def\ket{\rangle}
\def\Arg{\hbox {\rm Arg}}
\def\Re{\hbox {\rm Re}}
\def\Im{\hbox {\rm Im}}
\section{Introduction and Summary of Seiberg-Witten's Solution
for $N_f=1,2,3$;  Symmetry of the Models}

The celebrated works of Seiberg and Witten on $SU(2)$ gauge theories 
with $N=2$ supersymmetry\cite{SW1,SW2} have opened the way for 
exploring the low energy dynamics of  nontrivial four dimensional non 
Abelian gauge theories in a detailed and exact fashion (for 
generalizations see \cite{general}; for a review, see \cite{Rev}).

The  solutions presented by these authors indeed amount to the complete
determination of the vacuum degeneracy, to the exact calculation of the 
quantum (perturbative and nonperturbative) corrections to the low energy
effective couplings and $\theta$ parameter in each vacuum, and (with certain 
additional analysis \cite{BF}) to finding the exact specta of stable 
particles. These solutions exhibit interesting phenomena, such as the 
dynamical relation between confinement and chiral symmetry breaking, 
which could shed important light in the study of QCD.

In spite of the impressive breakthroughs accomplished in the original 
works of Seiberg and Witten and many works which followed, however, 
there seem to be still quite a few questions left unclear and which 
deserve further study. It is the purpose of this article to attempt to  
clarify (at least some of) those features, concerning in particular 
the structure of the low energy effective Lagrangian, CP properties of 
the theories and charge and quark number fractionalization associated 
to dyons, confinement and supersymmetry breaking.

The theories we are interested in are described  by the Lagrangian,
\beq
L= {1\over 8 \pi} \im \, \tau_{cl} \left[ \int d^4 \th \,
\Phi^{\dagger} e^V \Phi +\int d^2 \th\,{1\o 2} W W \right]
+ L^{(quarks)} + \Delta L +  \Delta^{'} L,   
\label{lagrangian}
\eeq
\beq
L^{(quarks)}= \sum_i [ \int d^4 \th \, \{ Q_i^{\dagger} e^V Q_i +
{\tilde Q_i}  e^{-V} {\tilde Q}_i^{\dagger} \} +  
\int d^2 \th \{ \sqrt{2} {\tilde Q}_i \Phi Q^i +  
m_i {\tilde Q}_i  Q^i \} + h.c.],
\eeq
where
\beq 
\tau_{cl}={\th_0 \o \pi} + {8 \pi i \o g_0^2}.
\label{struc}
\eeq
$\Phi= \phi + \sqrt2 \th \psi +\ldots,$ 
and 
$W_{\alpha} = -i \lambda + {i \o 2}(\si^{\mu} \sb^{\nu})_{\alpha}^{\beta} 
F_{\mu \nu} \th_{\beta} +\ldots $ 
are both in the adjoint representation of the gauge group (here taken to 
be $SU(2)$) while ($N=2$ quark hypermultiplets) $Q_i, {\tilde Q}_i$ are in 
the fundamental representation. In some part of the analysis below we 
consider also the addition of the mass term, 
\beq   
\Delta L=   \int \, d^2 \theta \,m_{A} \,\Tr \,
\Phi^2,
\label{adjointm}
\eeq
which reduces the supersymmetry to $N=1$, as well as supersymmetry breaking
terms such as
\beq
\Delta^{'} L =m^{'2}\,  \Tr \,\phi^2 +h.c..
\eeq

The main results of \cite{SW1,SW2} for the $N=2$ theories 
($ \Delta L= \Delta^{'}L =0$) 
may be summarized by the mass formula,
\footnote{We follow the convention of the second paper of Seiberg and 
Witten. In particular note the relative factor $2$ in the definition of 
$a$ and in $n_e$ with respect to those in their first paper. The mass of 
the adjoint field $\Phi$ will be denoted here as $m_A$  to distinguish 
it clearly from the quark masses $m_i$.}
\beq
{\cal M}_{n_m, n_e, S_i}=\sqrt{2} 
\left| n_m a_D + n_e a + \sum_i {m_i \o \sqrt{2}} \, S_i \right|, 
\label{massform}
\eeq
together with the curves (tori),
\footnote{The necessity of the shift in $u$ by a constant in the $N_f=3$ 
case as compared to the expression given in \cite{SW2} was found by 
Harano and Sato \cite{HS} by the explicit instanton calculations.}
\bea
y^2&=&x^2(x-u)+  {\Lambda_0^4 x \o 4};    \qquad (N_f=0), \non \\
y^2&=&x^2(x-u)+  { m_1 \Lambda_1^3 x \o 4}- {\Lambda_1^6 \o 64};
\qquad (N_f=1),\non \\
y^2&=&(x^2 - {\Lambda_2^4  \o 64})(x-u)+{m_1 m_2 \Lambda_2^2 x \o 4}
- {(m_1^2 + m_2^2) \Lambda_2^4 \o 64};    \qquad (N_f=2), \non \\
y^2&=&x^2(x-u-{\Lambda_3^2 \o 432}) -
{\Lambda_3^2 \o 64 }(x - u-{\Lambda_3^2 \o 432})^2 - {\Lambda_3^2\o 64 }
(m_1^2 + m_2^2 +m_3^2) \,\,
(x-u-{\Lambda_3^2 \o 432}) +   \non \\
& & + {\Lambda_3\o 4 }m_1 m_2 m_3  x -
{ \Lambda_3^2\o 64 }(m_1^2 m_2^2  +m_2^2 m_3^2 + m_3^2 m_1^2);
\qquad (N_f=3),
\label{curves}
\eea   
from which one finds
\beq   
{da_D \o du} =  {\sqrt{2} \o 8 \pi} \int_{\alpha} {dx \o y}, \qquad
{da \o du} =  {\sqrt{2} \o 8 \pi} \int_{\beta} {dx \o y},
\label{dadeda}
\eeq
as the two associated  periods. Finally, the vev $a$ or its dual $a_D$ 
themselves, can be obtained by integrating the above formula in $u$. 
$S_i $ represents the $i$-th quark number of the dyon under consideration.

The exact global symmetry of the models with quark hypermultiplets has been
studied carefully in Ref.\cite{SW2}.  For zero bare quark masses   it consists of
$SO(2N_f) \times SU_R(2) \times Z_{4(4-N_f)}$, where $Z_{4(4-N_f)}$
(which commutes with $N=1$ supersymmetry) transforms the fields as 
\bea     
& &\Phi \to   e^{2i \alpha} \Phi; \quad 
(Q^1, {\tilde Q}^1) \to e^{-i \alpha}({\tilde Q}^1, Q^1); \non \\
& & Q^{i} \to  e^{-i \alpha}  Q^{i}; 
\quad {\tilde Q}^i \to e^{-i \alpha}{\tilde Q}^i,  \quad (i \ge 2),
\label{Zsymm}
\eea
where $\alpha= 2\pi k/4(4-N_f)$ ($k=1,\ldots, 4(4-N_f)$).
The charges of various fields under the global $U(1)$ symmetries
are given in Table 1. Note also the relations
\beq 
Q_A=Q_R-2Q_J
\label{relation}
\eeq
and
\beq   
U_J(1) \subset SU_R(2), \quad
U_{\ell}(1) \subset SO(2 N_f). 
\eeq

\begin{table}{
\leftskip 2 cm
\rightskip 1cm
{\bf Table 1}: Global $U(1)$ symmetries of the models 
}
\begin{center}
\begin{tabular}{|c|c|c|c|c|c|c|c|c|c|}
\hline
Group &  Charge & $Q_i$ & $\psi_i$  &  ${\tilde Q}_i$ &
 ${\tilde \psi}_i$ &  $\phi$ & $\psi$ & $\lambda$& Comments \\
\hline
$U_R(1)$ & $Q_R$  & 0 & -1 & 0 & -1 & 2 & 1 & 1  & Anom. (A) \\
$U_A(1)$ & $Q_A$  & -1 & -1 & -1 & -1 & 2 & 2 & 0  & A.  \\
$U_J(1)$ & $Q_J$  & 1/2 & 0 & 1/2 & 0 & 0 & -1/2 & 1/2  & Non. Anom. (N A)  \\
$U_V(1)$ & $Q_V$  & 1 & 1 & -1 & -1 & 0 & 0 & 0  & N A\\
$U_{\ell}(1)$ & $S_{\ell}$  & $\delta_{\ell i}$ & $\delta_{\ell i}$ &
-$\delta_{\ell i}$ & -$\delta_{\ell i}$ & 0 & 0 & 0  & N A $\ell$-th
quark num.\\ \hline
\end{tabular}
\end{center}
\end{table}

\section{Number of the  Vacua of $N=1$ Theories}
\label{sec:numvac}

Before going into the analysis of the CP properties of the models, let us
discuss one aspect of the Seiberg-Witten solutions of $SU(2)$ gauge 
theories with $N_f=0,1,2, 3$, which is quite curious at first sight.  
Namely, the number of the singularities of the curves Eq.(\ref{curves})  
(singularities of the quantum moduli space), for unequal bare quark masses
$m_i's$, turns out to be
\beq 
{\cal N} = N_f +2. 
\label{numvac}
\eeq
For instance, for $N_f=2 $ and for small quark masses 
($|m_i| \ll \Lambda_2$), they are at 
\bea
u_{1,2} &=&  - {{\Lambda_2}^2\o 8} \pm {{\Lambda_2}\o 2}(m_1 + m_2)  - 
{ 10 m_1 m_2 + 3 m_1^2 + 3 m_2^2 \o 4} + O(m^3),  \non \\
u_{3,4} &=&  {{\Lambda_2}^2\o 8} \pm {{\Lambda_2}\o 2}(m_1-m_2) i + 
{ 10 m_1 m_2 - 3 m_1^2- 3 m_2^2 \o 4} + O(m^3).
\label{nf2sing}
\eea
What is the meaning of Eq.(\ref{numvac})?

A possible answer emerges from the observation that the  number
of singularities of quantum moduli space, i.e., the number of values 
of $u=\Tr \, \Phi^2$, at which some dyon becomes massless, is equal to the 
number of the vacua in  the theory with $N=1$ perturbation,
\beq   
\Delta S= \int d^4x \, d^2 \theta \,m_{A} \,\Tr \Phi^2
\label{perturb}
\eeq
(see the discussion Sec.\ref{sec:LEEF} below).
The latter, on the other hand, can be deduced directly from the scalar
potential of the original theory. In fact, for nonvanishing $m_A$ and 
nonvanishing and unequal quark masses $m_i$, the Lagrangian
Eq.(\ref{lagrangian}) leads to an $SU(2)$ invariant vacuum,
\beq    
\Phi= Q= {\tilde Q}=0, \label{symmet} 
\eeq
as well as $N_f$ isolated (classical) vacua.  For instance, for $N_f =2$
these "special points" are at
\footnote{The matrix  for the squarks is in the mixed color (column) and  
flavor (row) space.  Also, other nontrivial vacua are related to these
"special points"  by a color/flavor $SU(2)$ rotation.}
\beq  
\Phi= \pmatrix{a & 0 \cr 0 & -a}, \qquad 
Q^i_{k}= \pmatrix {a_1 & 0 \cr 0 & a_2}, \qquad
{\tilde Q}^{* i}_{k}= \pmatrix {{\tilde a}_1 & 0 \cr 0 & {\tilde a}_2},   
\eeq
with
\beq
a=-{m_1 \o \sqrt{2} }, \qquad  
a_1={\tilde a}_1=\sqrt{2 m_A m_1}, \qquad
a_2={\tilde a}_2=0,
\eeq
or
\beq
a={m_2 \o \sqrt{2} }, \qquad  
a_2={\tilde a}_2=\sqrt{2 m_A m_2}, \qquad
a_1={\tilde a}_1=0.  
\label{accid}
\eeq

For generic $N_f$, the vevs characterizing these  isolated vacua can be,
modulo color and global symmetry transformations, put in the form 
($k=1,\ldots N_f$)
\bea
& & \Phi^3 = 2 a  \quad (a=-{m_k \o \sqrt{2}});\quad
\Phi^{\pm} =0; \quad
 Q^i_{1} = 2\sqrt{m_A m_k}\,\delta_{ik}; \non \\
&& {\tilde Q}^{i}_{1}=  2\sqrt{m_A m_k}\, \delta_{ik}; \quad
 Q^i_{2} = {\tilde Q}^{i}_{2}= 0, \quad (i=1,2,\ldots,N_f).
\label{specialp}\eea
Such a result is best seen by first  going to the basis of quark fields 
which transforms as an $SO(2N_f)$ vector; see Appendix A.

Since the gauge symmetry is broken by the vevs of the squarks, in counting 
the number of vacua these special points contribute one each. Taking the 
multiplicity of the $SU(2)$ symmetric vacuum (\ref{symmet}) as $N_c=2$
(by  Witten's index argument \cite{index}), one thus sees that there are 
$N_f +2$ classical vacua.

If some mass is large compared to $\Lambda_{N_f}$, the theory correponding 
to the associated  special point is simply a QED like theory, with
a light (if $m_A \ll \Lambda$) electron (quark).

On the other hand, these $N_f +2$ vacua will mix in the infrared if  
$|m_i| \ll \Lambda_{N_f}$.  As long as supersymmetry remains exact, 
however, the number of the vacuum states - $N_f +2$ linearly independent 
states satisfying
\beq    
Q_{\alpha} |0_i\ket =0, \quad  
{\bar Q}_{\dot {\alpha}} |0_i\ket
=0,\quad
(i=1,2,\ldots N_f +2),
\eeq
where $Q_{\alpha}$ and  ${\bar Q}_{\dot {\alpha}}$  are supersymmetry
charges -  is expected to remain unaltered. We thus arrive at
Eq.(\ref{numvac}).

In this connection a quite nontrivial problem would be to reproduce the
vevs of
$u=\Tr \, \Phi^2$  by direct instanton calculations, in the manner of
\cite{AKMRV}, taking into account of the "accidental" degeneracies 
 Eq.(\ref{specialp}). 

This observation can also be interpreted as the confirmation of the
correctness of the counting of the number of vacua by Witten's index, 
at least for the $SU(2)$ 
(hence presumably also for $SU(N)$-) group.

In passing, an interesting recent observation by Kovner and Shifman
\cite{KS} seems to deserve a comment. Namely, it is suggested that 
$N=1$ supersymmetric pure Yang-Mills theory  with $SU(N_c) $ gauge group,   
might possess, beyond the standard $N_c$ vacua\cite{VY,AKMRV},  an extra vacuum with
\beq 
\bra \lambda \lambda\ket =0.
\eeq
Now, the massless sector of the original ($N_f=0$) $N=2$ Seiberg-Witten 
theory reduces, with a nonzero adjoint mass $m_A$, to that of the  
$N=1$ pure Yang-Mills theory discussed in \cite{KS}. The number of the 
vacua in this case has now rigorously been proven by Seiberg and 
Witten \cite{SW1} to be two:\footnote{
The original "conjecture" by Seiberg and Witten on the singularity
structure in the QMS hence of their exact solution,  has now been confirmed
by explicit instanton calculations.\cite{instantons,HS}
}
they are at  
\beq   
u= \bra \Tr \phi^2 \ket = \pm \Lambda_0^2, 
\eeq
namely at the two singularities of the QMS of the $N=2$ theory.
On the other hand,  the gaugino
condensate at these vacua is known from the anomaly of Ref \cite {KK}:
\beq    
\bra {g^2 \o 32 \pi^2 } \Tr \lambda \lambda \ket 
= 2 m_A u  =  \pm 2 m_A \Lambda_0^2. 
\eeq
The latter  reduces to $\pm \Lambda_{N=1}^2$ in the decoupling
limit, $ m_A \to \infty, \,\,  \Lambda_0 \to 0,$ with 
$\Lambda_{N=1}^2=m_A \Lambda_0^2$ fixed, reproducing the standard
Veneziano-Yankielovicz results \cite{VY,AKMRV}. These considerations cast
serious doubts on the validity of the conjecture of Ref \cite{KS}.

Note that in our argument an eventual  criticism on the difficulty of
controlling strong interaction dynamics \cite{KS} does not apply:  
it is precisely what 
Seiberg and Witten managed to bypass by use of duality and holomorphism.
Strictly speaking, our   reasoning  involves an  extrapolation of the adjoint
mass $m_A$ to values at least of the order of $\Lambda_0$; however, both the
counting of  the $N=1$ vacua \cite{SW1,SW2} (requiring the knowledge of the
superpotential only) and the anomaly of Ref \cite{KK},  are valid
independently of the value of $m_A$,  as long as $N=1$ supersymmetry is
unbroken.

\section{CP Invariance, Periodicity in $\theta$  and Electric Charge
Fractionalization} 
\label{sec:cp}

It is of  crucial importance to understand  the CP invariance properties of
the theory  to write down  correctly  the low energy effective Lagrangians
involving light monopoles and understand  their properties.  We first
discuss the electric charge of the dyons, leaving the discussion of  their
global quark numbers to the next  section.

\subsection{$m_i=0$ }
\label{sec:thetazerom}

First consider the theories with zero bare quark masses.  By the anomalous
$U_R(1)$ transformation any bare $\theta$ parameter can  be set to zero.
The theories are thus formally invariant under the CP transformation
\beq  
\phi \to \phi^{*}, \quad \psi \psi \to {\bar \psi} {\bar \psi},
\quad \psi_{qi} {\tilde \psi}_{qi} \to  {\bar \psi_{qi}} 
{\bar {\tilde \psi}_{qi}}
\eeq
together with the standard transformation of the gauge fields.
This  invariance  however is generally broken by the complex vev
$\bra \phi^{(3)} \ket \ne 0 $ \`a la Lee.\cite{Lee}  In terms of the 
gauge invariant order parameter $u= \bra \Tr \phi^2  \ket$  the CP 
violation is parametrized by $ \Arg \, u$.

In another convention in which $u$ is taken real and positive by an 
appropriate $U_R(1)$ transformation, the CP violation is seen as due to
the bare $\theta$ parameter
\beq 
\theta_0 = -{4-N_f \o 2} \, \Arg \, u.
\label{cpviol}  
\eeq

In the effective low energy theory at a generic point of QMS, CP violation 
of the theory is described by the effective theta parameter $\theta_{eff}$.
Somewhat surprisingly, the low energy effective theta parameter 
$\theta_{eff}$ computed from the exact Seiberg-Witten solution is found not 
to coincide, in the semiclassical limit, with  Eq.(\ref{cpviol}), but is 
related to the latter by (see below)
\beq   
\theta_{eff}\equiv  \Re {d a_D \o da}  \simeq
-{4-N_f \o 2} \, \Arg u -{\pi \o 2} N_f = \theta_0 -{\pi \o 2} N_f.
\label{shift}
\eeq
This appears to present us with a paradox. For example, for $N_f=1$,  at
large and real $u \simeq 2 a^2$ we expect the theory to be invariant 
under CP, while the exact Seiberg-Witten solution gives asymptotically
\beq 
\theta_{eff} =  -{\pi \o 2}, 
\eeq
as can be easily verified by using the explicit formulae given in
\cite{IY,BF}
\footnote{
Though $\theta_{eff}$ for $N_f = 2, 3$ found in \cite{IY,BF} is  
different from (\ref{shift}) by $\pi$, this difference does not affect
the physics because of the periodicity of $\pi$ as will be shown below. 
}. 
An analogous paradox appears for $N_f=3$. What is happening?

The key for solving  this apparent paradox is the fact that
the global symmetry of these models contains the discrete symmetry
$Z_{4(4-N_f)}$ (Eq.(\ref{Zsymm})), not $Z_{2(4-N_f)}$ na{\"{\i}}vely 
expected from the instanton argument. Such a transformation however 
induces the shift
\beq
\theta \to \theta + 2 (4-N_f) \beta = \theta + \pi,
\eeq
implying that physics is periodic in $\theta$ {\it with periodicity $\pi$}
unlike in ordinary gauge theories such as QCD or in the case of $N_f=0$.

The periodicity of  $\pi$ in the $\theta$ angle may sound somewhat 
surprising at first sight. It is not however the first time we encounter
nontrivial effects massless fermions exert  on the $\theta$ dependence of
a given theory: in the standard QCD,  massless fermions eliminate
the  vacuum parameter  dependence altogether, as is well known. The $\pi$
periodicity of the present theory is a combined consequence of the particular
interactions characteristic of $N=2$ supersymmetry  (which allow only the
particular set of $U(1)$ symmetries of Table 1) and of the $SU(2)$ nature 
of the gauge group.

The doubling of the anomaly-free discrete chiral symmetry itself, is present
in any theory such as QCD with $SU(2)$ gauge symmetry with quarks in the
fundamental representation. Consider in fact the nonvanishing chiral 
Green function  
($\psi = \psi_{L };\,\, {\tilde \psi} =  \psi_{R }^c$)
\beq
G= \bra T\{ \psi_{ i_1} {\tilde \psi}_{ j_1}
\psi_{i_2} {\tilde \psi}_{ j_2} \ldots \psi_{i_{N_f}} {\tilde \psi}_{
j_{N_f}} \}\ket
\label{thooft}
\eeq
(which corresponds to the nonzero, $SU(N_f)\times SU(N_f) $ symmetric
effective Lagrangian of 't Hooft)
is invariant under the discrete axial transformation with angle
$\alpha= 2 \pi k / 4 N_f$,  $k= 1, \ldots 4 N_f$
(under which $G$ either changes sign or remains invariant), combined with a
$Z_2$ transformation
$\psi_{1}  \leftrightarrow {\tilde \psi}_{1}$, $\psi_{i} \to {\psi}_{ i}$,
${\tilde \psi}_i \to {\tilde \psi}_i$, $i \ne 1$.  Note that such a
compensation of the minus sign in $G$ is not possible if all charged
particles are in the adjoint representaton of the gauge group. This 
explains the standard $2 \pi$ periodicity in $\theta$ found in the $N_f=0$ 
Seiberg-Witten theory \cite{SW1,DPK,Konishi}, whether or not
$N=2$ supersymmetry is broken to $N=1$ or to $N=0$.

Such a periodicity  means that the values of the $\theta$ parameter at
which theory is CP invariant,
includes half interger times $\pi$ such as $\pm \pi/2$. So the asymptotic
behavior,  $\theta_{eff} \to -\pi/2 $  as $ u \to +\infty $
mentioned above in the case of $N_f=1$,  is
perfectly consistent with the CP invariance the  theory is expected to have. It
remains however to explain the peculiar  shift of $\pi/2$.

This fact  turns out to be closely related to the phenomenon of charge
fractionalization of the magnetic monopoles. This problem
has recently been discussed in the present context by F. Ferrari.\cite{Ferrari}
In fact, the  semiclassical formula of
Ref \cite{Ferrari} (see Eq.(\ref{chargesc}) below) gives indeed the correct
shift of Eq.(\ref{shift}) in the massless case.

At or near the singularities of QMS, the effective low energy theory
involves light dyon fields.
See for instance Eq.(\ref{lagmono1}) below. These theories are strongly 
coupled at low energies. To select out such theories among the bare 
theories, we consider
adding the  adjoint mass term, $m_A \bra \Tr \Phi^2  \ket$.
Let us first  note that the phase of the
adjoint mass term  $m_A$  can  always be rotated away by the diagonal
subgroup of
$SU_R(2)$  as long as $N=1$ supersymmetry remains exact, hence is not
observable.

With zero bare quark masses the bare theory is  obviously invariant under CP,
since $\bra a\ket=0$ in the presence of  the adjoint mass.
On the other hand  the corresponding low energy effective theories
involving light
monopoles are characterized by various complex vev's;  the CP invariance
of these
effective theories is much  less  obvious.  However,  this can be
explicitly checked by generalizing the analysis made
in \cite{DPK} in the case of $N_f=0$ theory. The case of $N_f=1$ is
discussed in Appendix B.

Also, the singularities
for $N_f=1,2$ are related by the $Z_{4-N_f}$ symmetry of the
theory \cite{SW2}
\footnote{The discrete symmetry $Z_{4(4-N_f)}$ is
spontaneously broken
to $Z_{4-N_f}$ by the vev  $u$ which is invariant under $Z_4$.
}:
the low energy
$\theta$ parameter is accordingly shifted precisely by one period ($\pi$)
in going from one
singularity to another. This is another manifestation of the $\pi$
periodicity of the theory.

The value of $\theta_{eff}$ at the singularities
can in fact be easily determined as follows. At a singularity of the
quantum moduli space
where the  $(n_m, n_e)$ dyon
becomes massless (with $n_m \ne 0$), one finds from the exact solution that
\beq 
{n_m (d a_D / du) +  n_e (d a / du) \o (d a / du)}  =0. 
\label{theta0}
\eeq
It follows that
\beq 
\theta_{eff} = \Re {d a_D \o da}  \, \pi = -{n_e \o n_m}\pi
\label{theeff}
\eeq 
there,  hence the electric charge of such a  soliton is (by using the
formula of \cite{Witten}):
 \beq
{2 \o g}Q_e = n_e + {\theta_{eff} \o \pi} n_m =0.
\label{dyonismonop}
\eeq
The particle which becomes massless at a singularity (and which condenses
upon  $N=1$ supersymmetric perturbation  $m_A \Phi^2$ ) has always exactly
zero electric charge. It is a pure magnetic monopole, not a dyon.

Since, modulo monodromy transformations, 
$(n_m, n_e)=(1, 0)$, $(1, \pm 1)$ for $N_f=1$, 
$(n_m, n_e)=(1, 0)$, $(1, 1)$ for $N_f=2$ and 
$(n_m, n_e)=(1, 0)$, $(2, 1)$ for  $N_f=3$, 
for these singularities,  
the effective $\theta$ parameters take the values
\bea 
\theta_{eff} &=&  0, \mp \pi, \quad N_f=1; \non \\
\theta_{eff} &=&  0, -\pi,\quad N_f=2; \non \\
\theta_{eff} &=&  0, -{\pi \o 2},\quad N_f=3  
\label{valuesofth}\eea
there. 
For all $N_f$ these values of $\theta_{eff}$ are compatible with CP 
invariance of the bare theory.

\subsection{$m_i \ne 0$} 
\label{sec:thetanonz}

In the theories with non  zero bare quark masses, the situation is more 
subtle. First note that  (unlike in ordinary theories like  QCD or 
in $N=1$ supersymmetric QCD)
$\theta$ is  not the only CP violation parameter; the phases of $m_i$  
are another, independent source of CP violation. Since there is a unique
anomalous chiral $U(1)$ symmetry (see Table 1 and Eq.(\ref{relation})),
these phases cannot be absorbed into the $\theta$ parameter, once such a
$U(1)$ transformation has been  used to trade the phase of $u \sim 2a^2\,$ 
with $\,\theta_0$.

The central issue is  that when the theory is noninvariant under CP the
charge operator does not simply transform
under CP in the presence of monopoles \cite{JR,Wilczek}.
This fact lies under the  Witten's
formula \cite{Witten} for the electric charge of the monopoles
\beq  
{2 \o g}Q_e  = n_e + {\theta \o \pi } n_m.  
\label{Witten}
\eeq
In the $N=2$ theories under consideration
with more than one sources of CP violation  Witten's formula
gets generalized, and in the semiclassical limit, $|u| \sim |m_i|^2 \gg
\Lambda_{N_f}^2, $  Ferrari \cite{Ferrari} has found,   by using the earlier
calculation of Niemi, Paranjape and  Semenoff \cite{Niemi} that:
\beq
{2 \o g} Q_e =  n_e +
\left[ -{4 \o \pi}\Arg\, a + {1 \o 2 \pi}
\sum_{f=1}^{N_f} \Arg \, (m_f^2 -2 a^2)\, \right] \, n_m. 
\label{chargesc} 
\eeq

On the other hand, in the low energy effective action of Seiberg-Witten the
electric charge of a given $(n_m, n_e)$ particle should be  precisely given
by the Witten's formula Eq.(\ref{Witten}), with 
$\theta_{eff} =\Re (da_D/da) \, \pi$
in place of $\theta$, because this is the coupling strength of the monopole
with the electric field $A_{\mu}$ as can be deduced by use of the standard 
argument Ref \cite{Witten,Coleman}. It follows for consistency that
$\theta_{eff}$ must reduce, in the semiclassical limit, to the expression 
in the square bracket in the above formula (times $\pi$).
This can be explicitly checked by using the contour integration
representation for  ${da_D/du}$ and ${da/du}$. 

 In fact, one has  the exact solution\cite{SW2}
\bea 
{da_D \o du}&=&  { \sqrt{2} \o 8 \pi} \oint_{\alpha} {dx \o y},
\qquad
{d a \o du} \,\,= \,\,  { \sqrt{2} \o 8 \pi}
\oint_{\beta}{dx \o  y},
\label{contdadda}
\eea
where $\alpha$ and $\beta$ cycles  are appropriately
chosen so that $a_D$ and $a$  satisfy the  correct  boundary 
conditions in the semiclassical limit.  Using the contours specified 
in Appendix C
one finds  
 for general $N_f$ the result 
\beq
{d a_D \o da } = {i \o \pi}
\left\{ 
4 \log \, a - { 1\o 2} \sum_{i=1}^{N_f} \log  \, ({m_i^2 \o 2}-a^2 )+
\cdots
\right\},
\label{dadda} 
\eeq
hence
\beq
{\theta_{eff} \o \pi}=\Re {d a_D \o da }\simeq 
-{4 \o \pi}\Arg \, a + 
{1 \o 2\pi}\sum_{i=1}^{N_f} \Arg \, ( {m_i^2 \o 2} -a^2),
\label{thetasem}
\eeq 
in perfect agreement with Eq.(\ref{chargesc}).  
(Alternatively, one can use the asymptotic expansion found by
Ohta \cite{Ohta} by using the Picard-Fuchs equation, leading
to the same results.) 

A singularity where certain dyon becomes massless,  corresponds to a value of
$u$ at which one of the cycles (or a combination thereof) 
  collapses to a point. Now what matters for the
electric charge is $da_D/du$ and $da/du$.
And in the contour integrations defining them  (Eq.(\ref{dadeda})) neither the
problem of  residue nor that of integration constants (which are important in
the integrals defining  $a_D$ and $a$), are present.
As a result the argument  using
Eq.(\ref{theta0}), Eq.(\ref{theeff}) and Eq.(\ref{dyonismonop}) still holds 
in the cases with nonzero bare quark masses.  We conclude that the particle
which  condenses  upon  perturbation with $m_A \ne 0$ have always strictly 
zero electric charge: they are pure magnetic monopoles, 
although it does not imply CP invariance in general.

The periodicity in the $\theta$ parameter is also affected by the presence
of bare quark masses. One manifestation of this is that the $Z_{4-N_f}$ 
discrete symmetry acting on   QMS is explicitly broken by $m_i$.   
Nevertheless, physics is periodic in $\theta$ with period $(4-N_f)\pi$.

In the bare theory this can be seen by observing that  the mass terms (as
well as  the rest
of the Lagrangian) are invariant (see Table 1) under the discrete $U_A(1)$
transformations
with angles 
\beq     
\alpha = {\pi \o 2}, 
\label{transf}
\eeq
accompanied by the transformation
\beq  
Q_{i}  \leftrightarrow {\tilde Q}_{i}, \qquad \forall i. 
\eeq
The chiral transformation Eq.(\ref{transf}) corresponds to the shift of
$3 \pi$, $2 \pi$, and  $\pi$, for $N_f=1$, $N_f=2$ and  $N_f=3$, respectively.

In the effective theory such a periodicity can be seen as the invariance of the
theory under the monodromy transformation $u \to \exp(2 \pi i) \cdot u.$
From the monodromy at infinity, or more directly from Eq.(\ref{thetasem}),
one finds that $\theta_{eff}$ is shifted   precisely  by $-(4-N_f)\pi$.

\begin{figure}[hbt]
\begin{center}
\leavevmode
\epsfxsize= 8 cm
\epsffile{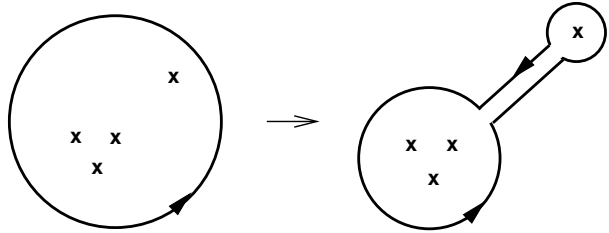}
\end{center}
\caption{
The modification of the contour upon decoupling of one fermion (e.g. from
$N_f=2$ to $N_f=1$).
$\times$ represents a singularity in the $u$-plane. }
\label{contour}
\end{figure}

Furthermore, the formula  Eq.(\ref{thetasem}) can  be  used to check the
consistency  between the $\pi$ periodicity of the massless theories and
$(4-N_f)\pi$ periodicity of massive theories, through decoupling. Namely,
consider sending one of the masses $m_f$ to infinity,  keeping 
$ \Lambda_{N_f-1}^2=(m_f)^{2/(5-N_f)}$ 
$(\Lambda_{N_f})^{2(4-N_f)/(5-N_f)}$
fixed and keeping the other $N_f-1$ quarks massless.
Clearly, in reducing the number of flavor by one  the monodromy at
"infinity" is modified by the monodromy around   $u \simeq m_f^2$ 
(see Fig.\ref{contour}).  The latter induces the shift of $-\pi$ as 
seen from  Eq.(\ref{thetasem}).
Thus in going from $N_f=3$ to $N_f=2$ one gets the periodicity in 
$\theta_{eff}$
\beq
-\pi - \pi = -2 \pi,
\eeq
but the massless $N_f=2$ theory has an exact $Z_2$ invariance, hence
the expected $\pi$ periodicity is recovered.   Analogously, in going from
$N_f=2$
to $N_f=1$, one gets  the apparent periodicity in $\theta_{eff}$
\beq   -2 \pi  - \pi = -3 \pi, \eeq
but due to the exact $Z_3$ invariance of the massless $N_f=1$ theory it
means the
true periodicity of $\pi$ again.
Finally, by going to $N_f=0$ one finds the
monodromy corresponding  to $\theta_{eff} \to \theta_{eff} -4\pi$, but because
of the $Z_2$ symmetry of that theory \cite{SW1}  the periodicity in
$\theta_{eff}$ is $2 \pi$ as expected.

From the above mentioned periodicity of $\theta_{eff}$ in the 
massive theories it follows that for generic values of $u$ and $m_i$, 
where the only light particles are the photon and its $N=2$ superpartners,
the low energy effective theory is CP invariant if $\theta_{eff} $ is an 
integer multiple of $(4-N_f) \pi/2$.

\section{Global Quark Numbers of Monopoles and Dyons}
\label{sec:qnum}

Another nontrivial manifestation of the lack  of CP invariance is the
fractionalization of  the quark numbers of the solitons.
The exact mass formula Eq.(\ref{massform}) of Seiberg and Witten for BPS
saturated states contains the ${i}$-th quark number $S_{i}$ of the
given $(n_m, n_e)$ dyon.   The knowledge of these quantum numbers is also
needed to write down the low energy effective Lagrangian involving light
monopoles, near the singularities of QMS.
First note however  that  the low energy effective
action (see Eq.(\ref{lagmono1}), Eq.(\ref{lagmono2}) and  Eq.(\ref{lagmono3}) 
below)
and the exact mass formula Eq.(\ref{massform}) are all  invariant under the
simultaneous shift
\bea    
a_D & \to &  a_D + \sum_{i=1}^{N_f}  c_i {m_i \o \sqrt{2}}; \non \\
S_i^{(n_m, n_e)}  &\to&  S_i^{(n_m, n_e)}  - c_i n_m, 
\label{shiftbis} 
\eea
where $c_i$'s  are arbitrary  real numbers.
At first sight such an  invariance  might appear to  prevent us from
determining unambiguously either $S_i^{(n_m, n_e)}$ or $a_D$.  
For the latter such an ambiguity would appear as an
indeterminancy of the integration constant
when the exact solution Eq.(\ref{dadeda}) is integrated in $u$.   
Note that the use of the exact mass formula fixes only the sum of such 
constants proportional to $m_i$'s appearing in Eq.(\ref{massform}).

This invariance reflects the fact that, even though the quark number 
associated to the $i$-th quark  (see Table 1) is a good
symmetry  (conserved and not spontaneously broken),
 there is no
way within the theory to measure it.\footnote
{This is analogous to the fact that
the electric charges of the ordinary quarks  are  conserved  by
strong interactions, but that within the latter there is no way to observe
their absolute values.
}  
This is to be contrasted with the case of the electric
charge of dyons discussed in the previous section.
Of course,
one could  make the $i$-th quark number  observable  by  gauging  weakly
the $U_{i}(1)$ symmetry but unless this is done any choice of  
$S_i$'s is as good as another.

Thus within the present theory one can fix $S_i$'s by some  convention
and accordingly determine the integration constants appearing in $a_D$.
(As was pointed out in \cite{Ferrari}, the arbitrariness of such a convention
should not be confused  with the ambiguity  of the contour defining  $a_D$,
as to whether or not  to pick up certain pole residues.   
The latter  gives rise to the
redefinitions of $a_D$ by integer multiples of $m_i/\sqrt{2}$ only
\cite{SW2}, hence
could at best  be regarded as a particular case of the mentioned 
arbitrariness of $S_i$.
Once $a_D(u)$ is defined by fixing the contour and the
integration constant, those residue contributions are however essential
to give the correct monodromy transformations, when $u$ is looped around a
singularity \cite{SW2}. )

For definiteness we take $S_i$'s to be the physical $i$-th
quark number of a given $(n_m, n_e)$ dyon, in the CP invariant theory
- the case in which $m_i $ and $u$  are all real and in the semiclassical 
region. 
This makes sense since the dependences on $m_i$'s are explicitly 
given in the exact solution of Seiberg-Witten (it appears in the form 
of the curves Eq.(\ref{curves}) and in the term proportional to $S_i$'s).
As is well known,  the quark numbers of the
magnetic monopoles are half integers  in a CP invariant theory\cite{JR}, 
see Table 2 and Appendix A.

Our $S_i$ can be thought as a possible choice for what was called
$s_i$ in \cite{Ferrari}, although  the meaning of the latter was 
somewhat obscure there.

\begin{table}{
\leftskip 1cm \rightskip 1cm
{\bf Table 2}: Quark numbers of the light dyons in the CP invariant theory;
$\pm$ denotes the  $SO(2N_f)$ chirality of the spinor representation.  
$\theta_{eff}$ gives the value of the effective $\theta$ parameter where the
corresponding  dyon
becomes massless. 
}
\bigskip \\
$N_f=1$
 \begin{center}
 \begin{tabular}{|c|c|c|c|c|c|}  \hline
 name &  S  & $n_m $ & $n_e$ & $SO(2)$ & $\theta_{eff}$ \\ \hline\hline
 $M$ & $-1/2$ & 1 & 0 & $+$ & $0$ \\ \hline
 $M'$ & $1/2 $ & 1 & 1 & $-$ & $-\pi$\\ \hline
 $M''$ & $-1/2$ & 1 & 2 & $+$ & $-2\pi$\\ \hline
 \end{tabular}
 \vskip .3cm
 \end{center}
$N_f=2$
 \begin{center}
 \begin{tabular}{|c|c|c|c|c|c|c|c| }  \hline
 name&  $S_1$  & $S_2$ & $n_m$ & $n_e$ &  $SO(4)$  & $SU(2)\times
 SU(2)$ & $\theta_{eff}$\\ \hline\hline
 $M_1 $ & $-1/2$ & $-1/2$ & 1 & 0 & $+$ 
 & (${\bf 2}, \,\, {\bf 1}$) & $0$\\
 $M_2$ & $1/2$ & $1/2$ & 1 & 0 & $+$ 
 & (${\bf 2}, \,\, {\bf 1}$) & $0$\\ \hline
 $M'_1 $ & $1/2$ & $-1/2$ & 1 & 1 & $-$ 
 & (${\bf 1}, \,\, {\bf 2}$) & $-\pi$\\
 $M'_2$ & $-1/2$ & $1/2$ & 1 & 1 & $-$ 
 & (${\bf 1},\,\,{\bf 2}$) & $-\pi$\\ \hline
 \end{tabular}
 \vskip .3cm
 \end{center}
$N_f=3$
 \begin{center}
 \begin{tabular}{|c|c|c|c|c|c|c|c|c|}  \hline
 name & $S_1$  & $S_2$ & $S_3$ & $n_m$ & $n_e$ & $SO(6)$ & $SU(4)$  
 & $\theta_{eff}$ \\ \hline\hline
 $M_0$ & $-1/2$ &$-1/2$ &$-1/2$ & 1& 0& $+$  & ${\bf 4}$ & $0$\\
 $M_1$ & $-1/2$ &$1/2$ &$1/2$ & 1& 0& $+$ &${\bf 4}$ & $0$\\
 $M_2$ & $1/2$ &$-1/2$ &$1/2$ & 1& 0& $+$ & ${\bf 4}$ & $0$\\
 $M_3$ & $1/2$ &$1/2$ &$-1/2$ & 1& 0& $+$ & ${\bf 4}$ & $0$\\ \hline
 $M'_0$ & $1/2$ &$1/2$ &$1/2$ & 1& 1& $-$  & ${\bf 4^*}$ & $-\pi$\\
 $M'_1$ & $1/2$ &$-1/2$ &$-1/2$ & 1& 1& $-$ &${\bf 4^*}$ & $-\pi$\\
 $M'_2$ & $-1/2$ &$1/2$ &$-1/2$ & 1& 1& $-$ & ${\bf 4^*}$ & $-\pi$\\
 $M'_3$ & $-1/2$ &$-1/2$ &$1/2$ & 1& 1& $-$ & ${\bf 4^*}$ & $-\pi$\\ \hline
 $N$ & $0$ &$0$ &$0$ & 2& 1&  & ${\bf 1}$ & $-\pi/2$ \\ \hline
 \end{tabular}
 \vskip .3cm
 \end{center}
\end{table}

Once the quark numbers $S_i$  of dyons are  fixed,   the integration constant  in 
$a_D$ is uniquely determined after specifying the integration contour
appropriately.     To determine it actually, one can for instance  start in a
semiclassical region ($u \gg \Lambda^2$, $m_i \gg \Lambda$), and takes  the
integration contour (the $\alpha$ cycle)  for instance  as in 
Appendix C.   The mass of a reference particle (say a $ (1,0)$  monopole)  should then
be computed with the exact mass formula Eq.(\ref{massform}) and we determine the
integration constant in $a_D$ such that it    coincides with the standard
semiclassical mass formula (see below, Eq.(\ref{massformsc})) in the appropriate
limit.   Alternatively, but equivalently, one can work directly in the strong coupling
region and  require that at the relevant singularity the reference particle 
  becomes
massless, according to the mass formula  Eq.(\ref{massform}).  For instance, the
$(1,0)$ monopole  should be massless at a singularity where the $\alpha$ cycle
collapses to a point.

Either way, we arrive at 
the formula
 \beq
a_D(u)=-{ \sqrt{2}\o 4 \pi}\oint_{\alpha}{dx \, y \o x^2 }
+ \sum_{i=1}^{N_f}{m_i \o 2 \sqrt{2}},
\label{adea} 
\eeq
where the $\alpha$ cycle is such that  it reduces to a zero cycle at one of the
$(1,0)$  singularities (corresponding to the massless monopole $M$ for $N_f=1$, to
$M_1$ for $N_f=2$, and  to $M_0$ for $N_f=3$, in Table 2).

As an example of a possible redefinition of the type,  Eq.(\ref{shiftbis}), one could
choose  the classical quark numbers    $S_i= S_i^{(cl)}$ (see
Eq.(\ref{clqnum})) instead of those given in Table 2. In such a convention,  $a_D$
would be given by the same contour integral as  in Eq.(\ref{adea}) but without  the 
constant term  proportional to $m_i$.

Note that, in contrast to the case of $a_D$,   $a$ does not suffer from the
arbitariness, Eq.(\ref{shiftbis}); accordingly the integration constant in the
case of $a(u)$ can be fixed uniquely by the  boundary  condition, 
$a(u)\sim\sqrt{2u}/2$ in  the semiclassical
region.  Alternatively, the 
integration constant for $a(u)$  may be fixed by the requirement 
that when a quark mass $m_i$ is large there must be a $(0,1)$  singularity 
at $a =\pm m_i /\sqrt{2}$.
For $N_f=1$, for instance,  this gives the expression
\beq
a(u) = -{ \sqrt{2} \o 4 \pi} \oint_{\beta}{dx \,y \o x^2 }
+{m \o  \sqrt{2}},   
\eeq
where the cycle $\beta$ is chosen so as to encircle the two smaller branch points.
See Appendix C for further discussions.

Of course,  in a CP {\it noninvariant} theory   the physical
$U_i(1)$  charges of the dyons differ from those
given in the Table 2: they are in  general fractional, not even half integers. In the
semiclassical limit  they can be computed in the standard manner; in the present
context,  Ferrari has obtained \cite{Ferrari}:
\beq
S_i^{(phys)} \simeq S_i^{(cl)}
+{n_m \o 2 \pi}\Arg {a + m_i \, /\sqrt{2} \o m_i /\sqrt{2} - a},
\label{qnumbsc}
\eeq
where $S_i^{(cl)}$ denotes  the integer classical (or canonical) $i$-th
quark number, Eq.(\ref{clqnum}).  For the semiclassical
dyons the classical charges are simply  the excitation numbers of 
fermion zero modes (see Appendix A).
The mass of a dyon is given semiclassically by $ \sqrt{2} |Z_{cl}|$ 
where
\beq   
Z_{cl} \simeq a \left({2 \o g}Q_e + i{8\pi \o g^2}n_m \right)
+ {1 \o \sqrt{2}}\sum_{i=1}^{N_f} \, m_i S_i^{(phys)}. 
\label{masssc}
\eeq
By using Eq.(\ref{chargesc}) and Eq.(\ref{qnumbsc}), 
this can be rewritten as
\bea    
Z_{cl} &\simeq& a n_e + 
(- {4 \o \pi} \Arg \, a + i{8 \pi \o g^2} )a\,n_m 
+{n_m \o 2 \pi} \sum_{i=1}^{N_f} \, 
\left[ ( a- {1 \o \sqrt{2} }m_i ) \,
\Arg \, ({m_i \o \sqrt{2}} - a) + \right. \non\\
& & + \left. (a+{1 \o \sqrt{2}}m_i)\,\Arg \,({m_i \o \sqrt{2}}+a)\right] 
+{1 \o \sqrt{2}}\sum_{i=1}^{N_f} \, m_i S_i^{(cl)}.
\label{massformsc} 
\eea

Again, such a result must be consistent with the exact Seiberg Witten 
solution in the appropriate, semiclassical limit.
The latter gives, as is seen easily
upon integration of Eq.(\ref{dadda}) with respect to $a$,
\bea
a_D \!\!&\simeq& \!\!
{i \o \pi}\left\{
4a\log a- {1 \o 2} 
\sum_{i=1}^{N_f} \,[\,
(a - {m_i \o \sqrt{2} })\log (a - {m_i \o \sqrt{2} })   + 
(a + {m_i \o \sqrt{2} })\log (a + {m_i \o \sqrt{2} })\, ] +
\right. \non \\
& & \left. + [\,(8-{N_f \o 2})\log 2 - (4 - N_f) \,] a -i{ N_f\pi \o 2}a
\right\} + \sum_{i=1}^{N_f}{m_i \o 2 \sqrt{2}}.
\eea
Inserted in the exact mass formula, Eq.(\ref{massform}), this is seen to reproduce
precisely the  electric and quark number charge fractionalization effects in the
semiclassical formula Eq.(\ref{massformsc}), besides  the one-loop renormalization 
corrections (which were not included in the latter).  This confirms the claim in Ref.
\cite{Ferrari} that the quark number  fractionalization effects reside in $a_D$, but
the main point of our result is that it provides a new,  far-from-trivial  check of  
the correctness of the  Seiberg Witten solution, Eq.(\ref{curves}) and 
Eq.(\ref{dadeda}).

As a by-product of this analysis, the above semiclassical mass
formula can be used to study  how the spectrum of various dyons
depends on the vacuum parameter. For simplicity we consider the case 
$m_i=0,$ $\forall i$. Indeed, in such a case the $\theta$ dependence 
appears in the semiclassical formula only through the factor, 
$|2 a Q_e /g_{eff}|^2= |n_e + (\theta_{eff}/\pi) n_m|^2$, with
$\theta_{eff} \simeq \theta_0 -{\pi \o 2} N_f$. 
The  value of $\theta_{eff}$ parameter where a given semiclassical
dyon takes a local minimum, is  shown in the last column of Table 2.
We note that the ``lightest dyon'' is always electrically neutral 
(monopole), and furthermore the values of $\theta_{eff}$ in Table 2 
is in agreement with those at the singularities, Eq.(\ref{valuesofth}); 
it is natural to conjecture that these lightest semiclassical dyons 
are smoothly connected along a straight path at fixed  
${\hbox {\rm Arg }}\, u$,  to the massless dyon at the singularity 
lying on the same ray.

\section{$N=1$ Perturbation and Low Energy Monopole Effective Action}
\label{sec:LEEF}

We are now in a position to write down and discuss the low energy effective
Lagrangian
near the singularities of QMS. To select out these
vacua we perturb  the theory with a small  adjoint mass term,
Eq.(\ref{adjointm}).
Furthermore,  we shall limit our discussion to the cases of small bare
quark masses
$m_i$;  if one of them is large, the associated singularity describes a
weakly coupled
QED like theory of  the quark (electron) \cite{SW2},  whose properties are quite standard.

Let us consider the cases $N_f=1$ and $N_f=2$ in this section, postponing the
discussion of the $N_f=3$  to the next subsection.

\subsection{Effective Lagrangian for $N_f=1$}

When the bare quark mass $m$ is small, the singularities are near
one of the points \cite{SW2} $u_{1,3}=\exp(\pm i \pi/3)$ and 
$u_2= -1$   (we use the unit  $ 3 \cdot 2^{-8/3}  \Lambda_1^2= 1$).   Let
us consider the theory near $u_3= \exp(-i\pi/3)$ where a $(1,0) $ monopole $M$ is
light.   The effective Lagrangian has the form,
\bea 
{\cal L}&=&{1\over 8\pi} Im
\,[ \int d^4\theta \, {(\de F(A_D)/\de A_D)} {\bar A_D}+\int d^2 \theta \,
{(\de^2 F(A_D)/ \de A_D^2)} W_D W_D/2 \,] \non \\
& &  + \int d^4\theta \,[ M^{\dagger}e^{V_D} M +
{\tilde M}^{\dagger}e^{-V_D}{\tilde M}]+
\int d^2 \theta \, {\cal P} + h.c.,  
\label{lagmono1}
\eea
where the superpotential is given by  ($S=-1/2$)
\beq  
{\cal P}=(\sqrt{2}A_D+S\,m){\tilde M}M + m_A U(A_D)
\eeq
and $A_D(u)$ ($U(A_D)$ is its inverse) is provided by the
Seiberg-Witten solution. By minimizing the superpotential one finds
\beq     
(\sqrt{2} A_D -{1 \o 2} m ) M =0, \quad
\sqrt{2}{\tilde M} M +   m_A U^{'}(A_D) =0,
\eeq
from which one gets the vacuum expectation values of these fields:
\beq   
\bra M \ket = \bra {\tilde M} \ket = 
\left(-{m_A U^{'}(A_D) \o \sqrt{2}}\right)^{1/2},\quad
a_D={m \o 2 \sqrt {2}}.  
\eeq
This shows that the continuous vacuum degeneracy of the $N=2$ theory is indeed
lifted and (locally) gives  the unique vacuum at $u_3= \exp(-i\pi/3)$.
As in \cite{SW1} the magnetic monopole condensation means confinement \cite{TM}, as
expected for a theory which become strongly coupled in the low energies.

Anologous situation presents itself if one starts  with the effective 
Lagrangian near $u_1$ or $u_2$: each singularity of the $N=2$ QMS leads 
to an $N=1$ vacuum.

\subsection {Effective Lagrangian for $N_f=2$}

The case of ${N_f=2} $ is  more interesting because of the presence of a
nontrivial chiral symmetry  $ SO(2N_f)= SO(4)$  (approximate if $m_i \ll
\Lambda_2$.)  There are now four singularities  of the QMS of the $N=2$ 
theory which become the vacua of the $N=1$ theory, by a by-now well 
understood mechanism through the modification of the superpotential 
produced by the adjoint mass term.
Let us consider the theory near two of the singularities $u_{1,2}$,
Eq.(\ref{nf2sing}), where   two $(1,0)$ dyons $M_1$, $M_2$ (see Table 2)   
are light. The effective Lagrangian describes those particles interacting
with the dual gauge multiplet $(A_D, W_D)$:
\bea 
{\cal L}&=&{1\over 8\pi} Im
\, [ \int d^4\theta \, {(\de F(A_D)/\de A_D)} {\bar A_D} +  \int d^2 \theta \,
{(\de^2 F(A_D)/ \de A_D^2)} W_D W_D/2 \,] \non \\
& & + \int d^4\theta \,\sum_{i=1}^2[ M_i^{\dagger}e^{V_D} M_i +
{\tilde M}_i^{\dagger}e^{-V_D} {\tilde M}_i] +
\int d^2 \theta \, {\cal P} + h.c.,  
\label{lagmono2}
\eea
where the superpotential is given by  (for $S_i$ see Table 2)
\beq  
{\cal P}=  \sum_{i=1}^2 ( \sqrt{2} A_D  +   \sum_{\ell=1}^2
S_i^{\ell}\, m_{\ell} ){\tilde M}_i M_i + m_A U(A_D).
\eeq
Supersymmetric vacua occur  where the superpotential is stationary:
\bea    
&& (\sqrt{2} A_D -{m_1 + m_2 \o 2}  ) M_1 =0,
\quad   M_1={\tilde M}_1,\non \\
&& (\sqrt{2} A_D + {m_1 + m_2 \o 2}  ) M_2 =0,
\quad   M_2={\tilde M}_2, \non \\
&&  \sqrt{2} \sum_{i=1}^2 {\tilde M}_i M_i +   m_A U^{'}(A_D)=0. 
\eea
These equations have two solutions,
\beq 
a_D= \bra  A_D \ket = {m_1 + m_2 \o 2 \sqrt{2}},
\quad M_2=0, \quad
\bra M_1 \ket  = \left(-{m_A U^{'}(A_D) \o \sqrt{2}}\right)^{1/2},
\label{vac1}
\eeq
and
\beq 
a_D= - {m_1 + m_2 \o 2 \sqrt{2}}, 
\quad M_1=0, \quad
\bra M_2 \ket  = \left(-{m_A U^{'}(A_D) \o \sqrt{2}}\right)^{1/2},
\label{vac2}
\eeq
corresponding respectively to the $u_1$ and  $u_2$ singularities of the 
$N=2$ theory. Note in particular the change of the value $\bra A_D \ket$ 
by $(m_1 + m_2)/\sqrt{2}$ in going from $u_1$ to the nearby singularity 
$u_2$, found here from the minimization of the effective low energy 
potential, is consistent with the behavior of the exact solution 
$a_D(u)$. In fact,  the two branch
points $x_1$ and  $x_2$ coincide both at  $u=u_1$ and at $u=u_2$; however,
in the
continuation the $\alpha$ cycle defining  $a_D(u)$ picks up twice
the contribution of the residue at  $x=-\Lambda_2^2/8$, which is precisely
$(m_1 + m_2)/\sqrt{2}$ as explained in \cite{SW2}.

Condensation of the monopole $M_1$ or $M_2$ leads to confinement
(\`a la 'tHooft-Mandelstam)  of the original color charges.    
But now the monopoles $(M_1, M_2)$  form a doublet 
({\bf 2}, {\bf 1}) with respect to the chiral 
$SU(2)\times SU(2)\sim {\widehat {SO(4)}}$.  The latter is therefore
spontaneously broken   to the group $SU(2)\times U(1).$  Moreover,
this  chiral symmetry breaking is a {\it consequence}  of the same
mechanism (monopole condensation) that is responsible for confinement.

We note also that in the $m_i \to 0$ limit (but with $m_A \ne
0$), the monopoles $M_2, \, {\tilde M}_2 $ (four real scalars) remain
strictly massless. They describe the three  Nambu-Goldstone bosons of 
the chiral symmetry breaking
$SU(2) \times SU(2)  \to SU(2)\times U(1) $ and two superpartners.
All other scalars contained in $A_D,\,\, M_1$  and  ${\tilde M}_1$ become
massive. (Their masses can be  analysed as in \cite{DPK}.)

In the strictly massless case ($m_1=m_2=0$)  the symmetry breaking
pattern is slightly different. Actually, the problem of finding the 
minima of the potential in this case reduces to that of finding the 
classical vacua  in the SQED with two flavors, discussed in
Sec. 2.5  of \cite{SW2}. The vacua has the general form (up to symmetry
transformations),
$M=(C, 0); \,\, {\tilde M}= -{\mu /\sqrt{2} C}, B),$
where  $\mu= m_A U^{'}(0)$,   and $B$ and $C$ are arbitrary complex numbers
satisfying $|C|^2=   |B|^2+  |C|^2/2|C|^2.$ The global symmetry is now 
broken as  $SU(2) \times SU(2)  \to SU(2).$

Though the pattern of the chiral symmetry breaking is at first sight
similar to what is believed to occur in the two-flavored QCD, the details 
are different, because of the
peculiar interactions special to the $N=2$ supersymmetry. In particular the
second $SU(2)$
factor in the $SU(2) \times SU(2)  \sim {\widehat {SO(4)}}$  symmetry is
not analogous to
the   axial $SU_A(2)$ in QCD  (the other $SU(2)$ is the standard vector
$SU_V(2)$.)
The properties of Nambu-Goldstone bosons (singlets of the remaining
$SU(2)$) are also
quite different from the case of QCD.

The theories near the other two singularities $u_{3,4}$ where the $(1,1)$ 
"dyon"
$M'_{1,2}$ are light, can be analysed in a similar manner. As explained  in
Sec \ref{sec:thetanonz}, the $(1,1)$ particles   are actually pure magnetic
monopoles with zero electric charge.

\section{Oblique Confinement at $N_f=3$}

The three ($N_f=1$) or  the two pairs of singularities ($N_f=2$) studied 
above are  physically quite similar when all bare quark masses are small: in
fact, in the limit of zero bare masses they are related by exact 
symmetries, $Z_3$ (for $N_f=1$) or $Z_2$ (for $N_f=2$).

In contrast, the singularity near
\footnote
{Recall $u^{'}= u + {\Lambda_3^2 / 432}.$ } 
$u^{'}= {\Lambda_3^2 / 256}$ at which the $(2,1)$
dyon becomes light, and the quartet of nearby singularities around
$u^{'}=0 $ associated with four $(1,0)$ monopoles, which occur in the 
$N_f=3$ case,  are physically very distinguished.    
Indeed, while the quartet of singularities are associated
with $(1,0)$  monopoles carrying flavor quantum number 
{\bf 4} of $SU(4) = {\widehat {SO(6)}}$ 
(see also the $S_i$ quantum numbers of $M_i$'s  in Table 2.),  
the singularity near $u^{'}= {\Lambda_3^2 / 256}$   
is due to a  $(2,1)$ dyon which is flavor neutral.

The analysis of the low energy effective Lagrangians is similar to 
the cases of $N_f=1$ or $N_f=2$. Near the quartet singularities, one has
\bea 
{\cal L}&=&{1\over 8\pi} Im
\,[\int d^4\theta \,{(\de F(A_D)/\de A_D)}{\bar A_D}+ \int d^2 \theta \,
{(\de^2 F(A_D)/ \de A_D^2)} W_D W_D/2 \, ] \non \\
& & + \int d^4\theta \,\sum_{i=0}^3 [ M_i^{\dagger}e^{V_D} M_i +
{\tilde M}_i^{\dagger}e^{-V_D} {\tilde M}_i ] 
+ \int d^2 \theta \, {\cal P} + h.c.,  
\label{lagmono3}
\eea
where the superpotential is given by  ($S_i$ in Table 2)
\beq  
{\cal P}=  \sum_{i=0}^3   ( \sqrt{2} A_D  +   \sum_{\ell=1}^3
S_{\ell}^{i} \,
m_{\ell} ){\tilde M}_i M_i +   m_A U(A_D),
\eeq
Minimization of the scalar potential leads to one of the four vacua:
\bea 
&&a_D= \bra  A_D \ket = {m_1 + m_2 + m_3 \o 2 \sqrt{2}}, \quad
M_i=0,\,(i\ne 0), \quad  \bra M_0 \ket  = (-{m_A U^{'}(A_D) \o
\sqrt{2}})^{1/2},\non \\
&& a_D={- m_1 - m_2 + m_3  \o 2 \sqrt{2}}, \quad M_i=0,\,(i\ne 1),  \quad
\bra M_1 \ket  = (-{m_A U^{'}(A_D) \o
\sqrt{2}})^{1/2},  \non \\
&& a_D={- m_1 + m_2 - m_3  \o 2 \sqrt{2}}, \quad M_i=0,\,(i\ne 2), \quad
\bra M_2 \ket  = (-{m_A U^{'}(A_D) \o
\sqrt{2}})^{1/2},  \non \\
&&  a_D=   { m_1 - m_2 - m_3  \o 2 \sqrt{2}}, \quad
M_i=0,\,(i\ne 3), \quad  \bra M_3 \ket  = (-{m_A U^{'}(A_D) \o
\sqrt{2}})^{1/2}.
\eea
In any one of them monopole condensation leads to color confinement
and at the same time, to dymanical chiral symmetry breaking, 
$SO(6) \to SU(3) \times U(1)$.
In the limit of  zero bare quark masses three of $(M_i, {\tilde M}_i)$
pairs remain massless ($12$ massless scalars); $6$ of them are the
Nambu-Goldstone bosons.

Again, in the case of strictly  zero bare masses (not the limit of massive
theory) some nontrivial vacuum degeneracy remains, and in a generic such 
vacuum, the symmetry is broken as   
$SO(6) \to SU(2) \times U(1)$.

All this has some similarity  to what happens in the
three-flavored  QCD at low energies, though the details are different
(for instance, the massless particles here are in the {\bf 3} 
of the remaining  $SU(3) $ symmetry, while the  pions in QCD are octets).

On the other hand, near the $(2,1)$ singularity, we have the low energy 
theory
\bea 
{\cal L}&=& {1\over 8\pi} Im
\, [ \int d^4\theta \, {(\de F(A_D)/\de A_D)} {\bar A_D} + 
\int d^2 \theta \, {(\de^2 F(A_D)/ \de A_D^2)} W_D W_D/2 \,] \non \\
& & + \int d^4\theta \,[ N^{\dagger}e^{V_D^{'}} N +
{\tilde N}^{\dagger}e^{-V_D^{'}} {\tilde N}] +
\int d^2 \theta \, {\cal P} + h.c.,  
\eea
where $A_D^{'}  \equiv 2 A_D + A$, $V_D^{'}  \equiv 2 V_D + V$ and   the
superpotential is given by   
\beq
{\cal P}=  \sqrt{2} A_D^{'} {\tilde N} N +   m_A U(A_D^{'}).
\eeq
Minimization of the scalar potential in this case leads to the condensation
of the $(2,1)$ magnetic monopole (confinement); however no chiral symmetry
breaking occurs in this vacuum, $N$ being flavor neutral.

This can be seen as an explicit realization of the oblique confinement of 't
Hooft. The dyons in this theory appear in a spinor representation of the
flavor group, $SO(6)\sim SU(4)$. In fact, the quartet  $M_i$ in
{\bf 4} of $SU(4)$  (Table 2) can be naturally identified, if
transported smoothly in  QMS into the semiclassical domain, with the
states
\beq 
|M_0\ket ; \quad |M_1\ket = b_2^{\dagger} b_3^{\dagger} |M_0\ket; \quad
|M_2\ket = b_3^{\dagger} b_1^{\dagger} |M_0 \ket; \quad
|M_3\ket = b_1^{\dagger} b_2^{\dagger} |M_0 \ket, 
\eeq
where $|M_0\ket$ is  the "ground state of the monopole sector",  
$b_i | M_0 \ket=0$. See Appendix A.

On the other hand, $N$ can be naturally interpreted as the 
bound state of the above $M_i$ states ({\bf 4} of $SU(4)$) 
and the $(1,1)$ states $M^{'}$ which belong to {\bf 4}$^*$ 
of $SU(4)$, constructed as
\beq 
|M^{'}_0\ket = b_1^{\dagger}b_2^{\dagger}
b_3^{\dagger} |M_0\ket;
\quad |M_1^{'}\ket = b_1^{\dagger} |M_0\ket;
\quad
|M_2^{'}\ket =  b_2^{\dagger} |M_0 \ket; \quad
|M_3^{'}\ket = b_3^{\dagger} |M_0 \ket.  
\eeq
Their quark numbers $(S_1, S_2, S_3)$ are given in Table 2, 
as is readily verified by using the formula, Eq.(\ref{qnumop}).
Note that the electric charges of the $M_i$  and  $M_i^{'}$ dyons in 
the vacuum near $u^{'}=\Lambda_3^2/256$ 
(where $\theta_{eff}=-\pi/2$, see  Eq.(\ref{theeff})) are
\beq   
{(2 / g_{eff})} Q_e(M_i)= -{1\o 2}, \qquad {(2 / g_{eff})} Q_e(M_i^{'})={1\o 2}; 
\eeq
we conjecture that their strong electrostatic attractions make their 
bound state much lighter than each of them. That would be precisely the 
phenomenon conjectured by t'Hooft \cite{TM} for QCD at $\theta=\pi$.

The existence of a classical solution corresponding to the $N$ state,
conjectured by Seiberg and Witten\cite{SW2} and later found in
\cite{2monopole}, does not contradict such an interpretation, 
since in the limit of large separations such a state is equivalent 
to the two-particle state made of $M_i$ and $M_i^{'}$.

One might wonder why oblique confinement occurs in the $N_f=3$ model,
while not for $N_f=1$ or $N_f=2$. Could not the singularity structure 
of the QMS in this case such that in one group of vacua 
(for small $m_i$) $M_i$'s are light, and in the other
$M_i^{*}$ are, such that in each vacuum there is ordinary confinement?
What is the criterion for ordinary or oblique confinement to be realized?

We have no general answer to these questions, though some remarks can be
made. To realize the above mentioned alternative dynamical 
possibility, the
theory with $N_f=3$ would need {\it eight} vacua at the ultraviolet cutoff,  
since the number of linearly independent states annihilated by supersymmetry 
generators is expected to remain invariant when the scale is varied  
from the ultraviolet to infrared. The bare theory in this case however 
possesses only $N_f+2=5$ vacua as was seen in Sec \ref{sec:numvac}; such
scenario is not allowed dynamically.  
The scenario realized in the exact Seiberg-Witten solution - oblique 
confinement  at $u^{'} \simeq {\Lambda_3^2 / 256} $  and  ordinary 
confinement near $u^{'} \simeq 0 $ (four vacua) - precisely matches with  
the five vacua present in the ultraviolet theory. 
Unfortunately, this explanation hinges crucially on the specific 
properties of supersymmetric theories, not easily generalizable to
ordinary theories.

\section{Soft Supersymmetry Breaking and $\theta$ Parameter Dependence
of Vacuum Energy Density}

As a further check that physics  discussed so far is reasonable, we 
study in this section the response of the system to a small 
perturbation which breaks supersymmetry. The analysis is a 
straightforward generalization of that made in \cite{Konishi}  (see also
\cite {Evans}) in 
the case of $N_f=0$, hence will be discussed only  briefly.
We start with the $N=1$ theory with small adjoint mass  $m_A$,
Eq.(\ref{adjointm}),  and for simplicity we set all bare quark masses to 
zero in this section.

As a simple  model of soft supersymmetry breaking we consider adding
\beq  
\Delta^{'} L =m^{'2} \Tr \Phi^2|_0 + h.c.            
\label{susybr}
\eeq
as perturbation.  ($|m^{'}| \ll |m_A| \ll |\Lambda|$.)
Since this theory  has a
finite number  of well defined, locally isolated  vacua,  the effect of the
supersymmetry breaking term Eq.(\ref{susybr}) on the energy density  can be
computed to first order, by using the standard perturbation theory:
\beq   
\Delta E = -\bra
\Delta^{'} L  \ket = - m^{'2} u + h.c. 
\label{energyd}
\eeq 
where $u$ is the
unperturbed  (i.e. $N=1$) vacuum expectation value which is known for each
vacuum. The
shift of energy density in general eliminates  the residual, $N_f+2-$ple
vacuum degeneracy. As the bare $\theta $ parameter
is varied, however, each of these local minima in turn
will play the role of the unique, true vacuum of the theory.

\begin{figure}[hbt]
\begin{center}
\leavevmode
\epsfxsize= 8cm
\epsffile{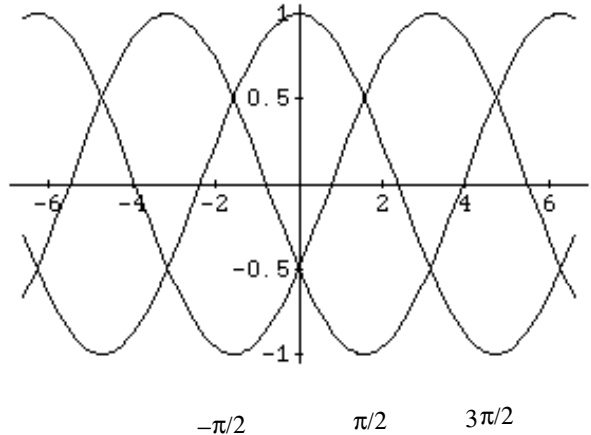}
\end{center}
\caption{Energy density in the three minima as a function
of $\theta_0^{'}$ }
\label{Cosine}
\end{figure}

We shall illustrate the situation for $N_f=1$.  In the three $N=1$ vacua
$u$ takes the values 
\beq    
u_1=e^{\pi i /3}A, \quad u_2= -A, \quad    u_3=  e^{-\pi i /3}A, 
\eeq
where $A\equiv 3 \cdot 2^{-8/3}\Lambda_1^2$.
The phase of $A$ is found by the observation that for general
$N_f $ ($N_f=1,2,3 $), $\Lambda_{N_f}$ depends on $\theta_0$ and on
$g$ through the combination,
\beq    
\Lambda_{N_f} = \mu \, e^{ \pi \tau i/ (4-N_f) },
 \quad   \tau= {\theta \o \pi}
+ { 8 \pi i \o g^2}.  
\eeq
For $N_f=1$ it means that $A \propto \exp{ 2 i \theta_0  /3}$. 
Therefore  one finds from Eq.(\ref{energyd}) the dependence of the energy
density in the three local minima on $\theta_0$:
\bea    
& &\Delta E = -2 |m^{'2}A|\,\cos ({2 \theta_0 \o 3 }+
{\pi \o 3} + 2 \Arg m^{'}), \quad  u\simeq u_1, \non \\
& &\Delta E = - 2 |m^{'2}A|\,\cos ({2 \theta_0 \o 3 } +
{\pi} + 2 \Arg m^{'}), \quad  u\simeq u_2, \non \\
& &\Delta E = - 2 |m^{'2}A|\, \cos ({2 \theta_0 \o 3 } -
{\pi \o 3} + 2 \Arg m^{'}), \quad  u\simeq u_3. 
\eea
These are illustrated in Fig.\ref{Cosine} where the vertical scale is
arbitrary.

The true vacuum  is thus near   $u_1$, $u_3$ and $u_2$,
for the value of $\theta_0^{'}\equiv \theta_0 + 3 \Arg m^{'}$
which lie in the regions,   $ -\pi < \theta_0^{'} < 0;$
$ 0 < \theta_0^{'} < \pi;$  $ \pi < \theta_0^{'} < 2 \pi,$
respectively.
Note that, in spite of the appearance of the cosine of angle
${2 \theta_0 /  3}$, the energy density of the true vacuum is a 
periodic function of $ \theta_0$ with period  $\pi$, in accordance
with the general discussion of Sec. \ref{sec:thetazerom}
At $ \theta_0^{'} = n \pi,$ $n=$ an integer, the vacuum is doubly 
degenerate, CP is spontaneously broken \`a la Dashen \cite{Dashen}.
As long as $m^{'}$ is small, the vev $u$ remains near one of
the $N=2$ singularities (at $ \theta_0^{'} = n \pi$  it jumps from one
local minimum to another), meaning that $\theta_{eff} \ll 1,$  
independently of the value of $\theta_0:$  CP is only slightly broken. 
The $M$ particle which condenses is a true dyon with small nonzero 
electric charge.\cite{Konishi}
Confinement persists at  nonzero  $\theta_0$, in contrast to the 
conjecture of \cite{Schier} but in accord with \cite{Cardy}.

\begin{figure}[hbt]
\begin{center}
\leavevmode
\epsfxsize=0.7\textwidth 
\epsffile{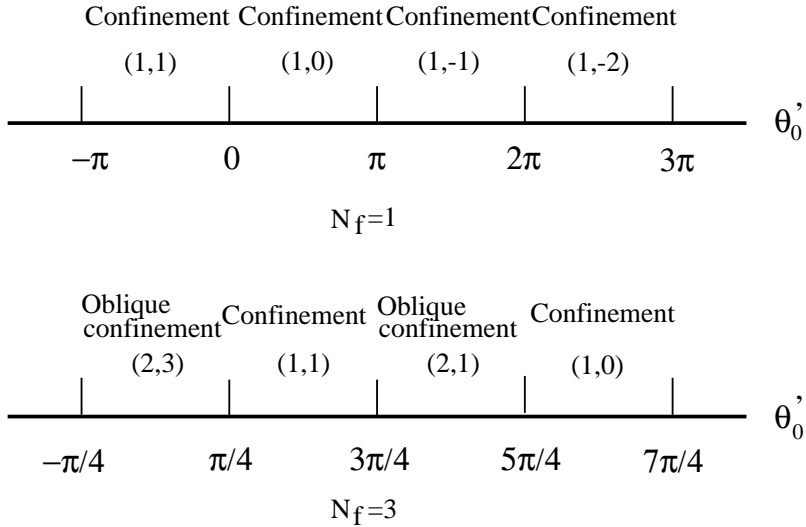}
\end{center}
\caption{
The $\theta'_0$ dependence of the phase structure in the cases 
$N_f=1$ and $N_f=3$.
}
\label{phase}
\end{figure}

The analysis can be repeated for $N_f=2$ or $N_f=3$: in all massless 
cases, the periodicity in $\theta_0$ of the energy density of the 
vacuum is found to be $\pi$ reflecting  the $Z_{4-N_f}$ symmetry in the 
$N=2$ theories. 
The phase diagram in the cases $N_f=1 $ and $N_f=3$ are shown in 
Fig.\ref{phase}.
In the massive cases, the discrete symmetry $Z_{4-N_f}$ is explicitly
broken: the periodicity in $\theta_0$ is 
reduced to $(4-N_f)\pi$. 

\section{Vacuum Structure of  the $N=1$ Effective Theories
involving "Meson" Fields}

In previous sections we examined the $\theta$ angle dependence
of the low energy effective theories involving magnetic monopoles, 
in the case the adjoint mass $m_A$ was small. 
(In fact, these effective actions are smoothly connected to the 
$N=2$ actions in the $m_A \to 0 $ limit.)  For large adjoint
masses, however, these effective theories are not the proper low-energy
descriptions of the theory. The more adequate low energy theory in 
those cases should involve "meson" variables  
$V^{IJ}=-V^{JI}=\hat{Q}^I\hat{Q}^J$ instead \cite{SW2,sei,elitzur},
and contain the superpotential $(1/m_A) (V_{IJ})^2$
obtained after the integration of the heavy adjoint fields.  
In the limit,
\beq 
m_A \rightarrow \infty; \quad
\tilde{\Lambda}_{N_f}^{6-N_f}=m_A^2\Lambda_{N_f}^{4-N_f}
\quad {\hbox{\rm fixed}}, 
\label{n1limit}
\eeq
these theories must smoothly be connected to the ordinary N=1 $SU(2)$ 
SQCD with $N_f$ flavors.

For intermediate values of $m_A$, we have thus two equivalent descriptions
of vacua \cite{SW2}; one involving the meson fields, and the other in 
terms of the photon and monopole fields. The number of vacua  and the 
$\theta$ dependence following from the low energy description involving 
meson fields must be the same as those found in Sec.\ref{sec:cp}  On the 
other hand, these properties should also match those of the N=1 SQCD 
vacua in the limit, Eq.(\ref{n1limit}).

For example, for  $N_f=1$  the only independent composite field
is  $V^{12}=iQ\tilde{Q}=i T$. The exact superpotential \cite{sei} 
is given by
\beq
W_{N_f=1}=\frac{m_A^2 \Lambda_1^3}{T} - \frac{1}{m_A}T^2 + mT.
\eeq
First  consider the massless case ($m=0$). The vev of T is given
by the solution of $T^3=-m_A^3\Lambda_1^3/2$:
there are three distinct vacua, just as in the $N_f=1,$ $N=2$ theory.
In fact, these three vacua are related by the phase rotation of
$T= Q\tilde{Q}$ field by $e^{i \pi/3}$ which is equivalent to the 
shift of $\theta$ by $\pi$, in agreement with the $\pi$ periodicity 
of the underlying theory.

For nonzero bare quark mass $m$, there are still three vacua, however,
they are no longer related by a $Z_3$ symmetry in $U(1)_R$ transformation.
Therefore the periodicity in $\theta$  is reduced to $3\pi$, again in 
agreement with the discussion of Sec.\ref{sec:thetanonz}

However when the adjoint fields are truly decoupled in the limit,
Eq.(\ref{n1limit}),  the $U(1)_A$ transformation becomes an independent 
chiral transformation (see Table 1).  As a result the periodicity in 
$\theta$ gets back to $\pi$, which is the correct periondicity of the 
massive $SU(2)$ SQCD.
\footnote{
The fact that physics is periodic in $\theta$ with periodicity $\pi$ for 
any $N_f$ in massive SQCD with $SU(2)$ gauge group, can be seen by using 
an argument similar to those used in Sec \ref{sec:cp}.}

In the $N=1$ limit, Eq.(\ref{n1limit}), we indeed recover all the known 
results of $N=1$ SQCD with a single flavor: all vacua "run away" 
($T \to \infty$) in the $m \to 0$ limit. On the other hand, if $m \ne 0$ 
there remain two vacua with finite vev $T_1, T_2$ (in accordance with the 
Witten's index for a massive $SU(2)$ gauge theory), though the vacuum
with $T_3$  runs away:
\beq
T_{1,2} \sim \pm \left({m_A^2 \Lambda_1^3 \o 2m}\right)^{1/2}, \qquad
T_{3} \sim 2 m m_A \,\rightarrow\,\infty.
\eeq 
Here the running solution is just the ``special point''
found through the classical analysis done in Sec \ref{sec:numvac}

A similar check of the number of vacua as well as  the $\theta$ angle
dependence in the low energy $N=1$ effective theories with meson 
degrees of freedom, can be made in the cases $N_f=2$ and $N_f=3$, 
using the superpotential \cite{sei},
\bea
W_{N_f=2}&=&
X(\mbox{Pf} V - m_A^2\Lambda_2^2) + W_{\mbox{tree}}, \non\\
W_{N_f=3}&=&
-\frac{\mbox{Pf} V}{m_A^2\Lambda_3} + W_{\mbox{tree}},
\eea
where
\beq
W_{\mbox{tree}} = \frac{1}{8m_A} \mbox{Tr} V^2
+\sum_{i=1}^{N_f} m_i V^{2i-1,2i}.
\eeq
In all cases we find the correct $\theta$ periodicity,  which is 
$\pi$ (zero bare quark masses), or  $(4-N_f)\pi$ (nonzero bare quark 
masses). Also, in the massless case one finds the continuous vacuum 
degeneracy, corresponding to the "Higgs" branch of vacua of the 
underlying $N=2$ theory.   In the massive cases the Higgs branches
disappear, instead there remain the $N_f+2$ distinct vacua. Again $N_f$ 
of them correspond to the "special points", and  run away to infinity 
in the $N=1$ limit.

\section{Conclusion}

In this paper we  explored several dynamical aspects of the $SU(2)$
Seiberg-Witten models with quark hypermultiplets, with a hope  to 
learn more about the detailed behaviors of a strongly interacting 
non Abelian gauge theories in four dimensions with fermions. Among 
others, we have clarified the meaning of the number of $N=1$ vacua
(number of singularities of QMS of the $N=2$ theories) as well as 
the CP properties and $\theta$ angle dependence of the theories with 
and without bare quark masses. We have shown in particular that the 
electric charge and quark number fractionalization of dyons is
correctly incorporated in the Seiberg-Witten solution, providing a new, 
strong confirmation of their beautiful results.

The structure of $N=1$ low energy effective Lagrangians and the 
correlation between the phenomenon of confinement and chiral symmetry 
breaking are also very interesting. Both dynamical possibilities 
(i.e., confinement either implies or not chiral symmetry breaking) 
are realized  in various $N=1$ vacua in these models. There seems to be
a deep and intriguing relation between the low energy phenomena 
(confinement or oblique confinement; chiral symmetry breaking) 
and the structure of classical vacua, though such a stringent UV-IR 
connection can be seen here only thanks to special properties of 
supersymmetris theories. 

There remains the task of applying what has been learned here to a more 
realistic field theory model of strong interactions: QCD. 

\bigskip
\noindent{\bf Acknowledgments}

The authors acknowledge useful discussions with Andrea Capelli and 
Camillo Imbimbo.  
One of us (H.T.) thanks the Dipartimento di Fisica dell'Uni\-ver\-sit\`a 
di Genova as well as the Genova section of INFN for a warm 
hospitality. He is also grateful for the financial support by the 
CNR-JSPS exchange program.

\appendix
\section{Flavor Quantum Numbers of  Magnetic Monopoles}

In the presence of  $N_f$ quark hypermultiplets belonging to the
fundamental representation of the $SU(2)$ gauge group, the flavor 
symmetry is enhanced to $SO(2N_f)$ (not to $O(2N_f)$ due to anomaly) 
due to the equivalence of {\bf 2} and {\bf 2}$^*$. The quark chiral 
fields $Q_a^i$ and $\tilde{Q}_i^a$ ($i=1,...,N_f$), which are in 
{\bf 2} and {\bf 2}$^*$, may be rearranged to form a set of $2N_f$ 
chiral fields in {\bf 2} as $Q_a^{2i-1}=Q_a^1$ and
$Q_a^{2i}=\epsilon_{ab}\tilde{Q}_i^b$ ($i=1,...,N_f$).
To see the $SO(2N_f)$ symmetry of the present model Eq.(\ref{lagrangian}),
however, we must consider an  $SO(2N_f)$ vector  $\hat{Q}_a^I$ formed  by
\beq
\hat{Q}_a^{2i-1}=\frac{1}{\sqrt{2}}(Q_a^{2i-1}+Q_a^{2i}), \quad
\hat{Q}_a^{2i}=\frac{1}{i\sqrt{2}}(Q_a^{2i-1}-Q_a^{2i}), \qquad
(i=1,2,\ldots N_f).
\eeq

It is well known\cite{JR} that each isospinor Dirac fermion $\psi_D^i(x)$ 
($i=1,...,N_f$) posesses one charge self-conjugate zero mode $u_0(x)$ in 
the monopole background. In the semiclassical quantization we expand 
$\psi_D^i(x)$ accordingly as
\bea
\psi_D^i(x)=b^i u_0(x) + \mbox{(non-zero modes)},
\nonumber \\
{^{\cal C}\psi_D^i}(x)=b^{i \dagger} u_0(x) + \mbox{(non-zero modes)},
\eea
where the suffix ${\cal C}$ denotes the charge conjugation. The canonical
commutation relation requires $\{b^i, b^{j \dagger}\}=\delta^{ij}$. 
The $2^{N_f}$ monopole states related by the action of these fermion 
zero-energy-mode operators are all degenerate. These states form a spinor 
representation of $SO(2N_f)$ group: the gamma matrices satisfying the 
$2N_f$ dimensional Clifford algebra are given by
\beq   
\gamma^{2i-1}= b^i + b^{i \dagger}, \quad
\gamma^{2i}=(b^i - b^{i \dagger})/i, \qquad i=1,2,\ldots  N_f, 
\eeq
and in terms of these the generators of the $SO(2N_f)$ group are given by:
\beq   
\Sigma_{ij}={1 \o 4i}[\gamma_i, \gamma_j], \quad   i \ne j,
\eeq
where $i, j = 1,\ldots, 2N_f$. Furthermore, the states with even (odd) 
number excitations belong to a chiral (anti-chiral) spinor representation,
since the $SO(2N_f)$ chirality operator is defined as
\beq
\gamma^{2N_f+1} =(-i)^{N_f} \gamma^1 \gamma^2 \cdots \gamma^{2N_f}
=\prod_{i=1}^{N_f} (1 - 2b^{i \dagger}b^i). 
\eeq

The $i$-th quark number charges of monopoles $S_i$ ($i=1,...,N_f$), just
correspond to the Cartan subalgebra of $SO(2N_f)$;
$\Sigma_{12},\, \Sigma_{34},\,\ldots, \Sigma_{2N_f-1,2N_f},\,$
since in a CP invariant theory they are given (in the subspace of the 
degenerate monopoles) by
\beq 
S_i=\int d^3 x \, {1 \o 2}({^{\cal C}\psi_D^i}\psi_D^i
- \psi_D^i{^{\cal C}\psi_D^i} )
=b^{i \dagger}b^i - 1/2 = \Sigma_{2i-1, 2i}.   
\label{qnumop}
\eeq
These charges take values $1/2$ or $-1/2$. We can define also the classical
(or canonical) quark number charges of monopoles, given 
by 
\beq  
S_i^{(cl)}=b^{i \dagger}b^i,   
\label{clqnum}\eeq
which are integers.

The generators of the spin group can also be easily constructed from 
the above; for instance for $N_f=2$, the two $SU(2)$ generators can be 
taken as
\beq   
T_1^{\pm } = -{i \o 2} ( \Sigma_{23} \pm  \Sigma_{41} ), \quad
T_2^{\pm } = -{i \o 2} ( \Sigma_{31} \pm  \Sigma_{42} ), \quad
T_3^{\pm } = -{i \o 2} ( \Sigma_{12} \pm  \Sigma_{43} ). 
\eeq

The important  point is that the values of the quark numbers $S_i$ and other
global quantum
numbers  taken by each member of a   monopole multiplet  belonging to
a particular
spinor representation,  are uniquely determined,  by using  these 
semiclassical constructions. Once such a connection is established, 
those values can be used  for
generic monopole states which are not necessarily semiclassical, and can even become
massless.  The
entries of Table 2 have been determined in this manner.

\section{CP Invariance of the $N_f=1$ Theory with $m_A\ne 0$, $m=0$}

In this Appendix we verify the CP invariance of the effective low energy theory
for  $N_f=1,$  $m_A\ne 0$ and  $m=0 $ described by
\bea 
{\cal L}&=&{1\over 8\pi} Im
\, [ \int d^4\theta \, {(\de F(A_D)/\de A_D)} {\bar A_D} + \int d^2 \theta \,
{(\de^2 F(A_D)/ \de A_D^2)} W_D W_D/2 \,] \non \\
& & +\int d^4\theta \,[ M^{\dagger}e^{V_D} M +
{\tilde M}^{\dagger}e^{-V_D} {\tilde M}] +
\int d^2 \theta \, {\cal P} + h.c., \eea
\beq  {\cal P}=  \sqrt{2} A_D {\tilde
M} M +   m_A U(A_D).
\eeq
Expanding all scalar fields around their vevs, (we consider for
definiteness the vacuum $u=u_3$),
\beq
a_D=0, \qquad
\bra M \ket = \bra {\tilde M} \ket = 
\left(-{m_A U^{'}(0) \o \sqrt{2}}\right)^{1/2}, 
\eeq
one sees that nontrivial phases appear in the fermion mass terms and in the
(generalised) Yukawa interactions,
\bea   
L_{Y}&=&\sqrt2 [ - \bra M^{\dagger} \ket \lambda_D (\psi_M
-\psi_{\tilde M}) + \bra M \ket \psi_D ( \psi_{\tilde M}  + \psi_M) ]
+{m \over 2} U^{''}\!(0) \psi_D  \psi_D ] \non \\
& & + \sqrt2 [ - M^{\prime \dagger } \lambda_D \psi_M +
{\tilde M}^{\prime  \dagger} \lambda_D \psi_{\tilde M} -
(A_D \psi_M \psi_{\tilde M} + \psi_D \psi_{\tilde M} M^{'} +  \psi_D
\psi_M {\tilde M}^{'})]\non \\
& & + {m \over 2} (U^{''}\!(A_D)- U^{''}\!(0)) \psi_D  \psi_D
\,\,+\,\, h.c. 
\label{Yukawa} 
\eea
where we introduced 
$M=\bra M \ket + M^{'}, {\tilde M}=\bra {\tilde M} \ket + {\tilde M}^{'}$. 
To see the phases appearing in  $U^{'}(0)$ and in $U^{''}(A_D),$  one needs 
the inverse of the Seiberg-Witten solution \cite{SW2,BF}
\beq   
a_D(u)={1 \o 6 \sqrt{2}}e^{-2 \pi i/3}(u^3 +1) 
F\left({5 \o 6}, {5 \o  6}; 2; 1 + u^3 \right). 
\label{hyperg}
\eeq
To invert this near $u= u_3= \exp(-i\pi/3)$ (in this Appendix we use the
unit $3\cdot 2^{-8/3} \Lambda_1^2=1$ used in \cite{BF}), first change the 
variable from $u$ to $w$ by
\beq   
u =u_3 (1 + w ), 
\eeq
so that
\beq   
u^3 +1 = u_3^3 (w^3 + 3 w^2 + 3 w) = - (w^3 + 3 w^2 + 3 w).
\eeq
Since the hypergeometric function is analytic around the origin with real
expansion coefficients, this
means that  $U(A_D)$ has an expansion of the form
\beq 
U(A_D) = u_3\{ 1 + w(e^{2 \pi i/3} A_D)\}, 
\eeq
where $w(x)$ is a power series in $x$ with real coefficients.

The phases appearing in Eq.(\ref{Yukawa}) can now be made explicit:
\bea 
&& \bra M \ket = (-{m_A U^{'}(0)
\o \sqrt{2}})^{1/2}=| \bra M \ket | e^{{i \o 2} \beta  + {\pi i \o 6}};
\quad \beta \equiv \Arg \, m_A\non \\ 
&&   m_A U^{''}(A_D)= -m_A \, w^{''}(e^{2 \pi i/3} A_D). 
\eea

At this point it is a matter of direct check to show that the phase 
rotations by $U(1)_R$ and $U(1)_J$ (see Table 1)
\bea 
&&  A_D \to e^{- 2 \pi i/3} A_D, \quad 
\psi_D \to,e^{-{i \o 2} \beta  - {\pi i \o,2}} \psi_D, \non \\
&& M^{'}  \to e^{ {i \o 2}\beta  +  {\pi i \o,6}}  M^{'}, 
\quad \psi_M \to e^{{\pi i\o 3}} \psi_M; \non \\
&&\lambda_D \to e^{ {i \o 2} \beta - {\pi i \o,6 } } \lambda_D. 
\eea
(the transformations of ${\tilde M}^{'}$ and $\psi_{\tilde M}$ are similar to
those of $M$ and $\psi_M$), eliminate all the phases from the Lagrangian
Eq.(\ref{Yukawa}):  the theory is indeed invariant under CP.

\section{Contour-integral representations for
$a_D(u)$ and $a(u)$; semiclassical behavior of $\tau$}

First consider the integrals 
\beq
{da_D \o du}\,=\,{ \sqrt{2} \o 8 \pi} \oint_{\alpha} {dx \o y},\qquad
{d a \o du}\,=\,{ \sqrt{2} \o 8 \pi} \oint_{\beta}{dx \o  y},
\label{contdaddabis}
\eeq
where $y$ is given in the form of 
$y = [(x-x_1)(x-x_2)(x-x_3)]^{1/2}$ and 
$\alpha$ and $\beta$ cycles encircling two branch points of $y$
are appropriately chosen as will be specified below.
Since these integrals  can be expressed in terms of the hypergeometric 
functions
\bea 
{da_D \o du} = i {\sqrt{2}\o 4}(x_2-x_1)^{1/2} 
F\left({1 \o 2}, {1\o 2}; 1; {x_2-x_3 \o x_2-x_1}\right), \non \\
{da \o du} = i {\sqrt{2}\o 4}(x_2-x_3)^{1/2} 
F\left({1 \o 2}, {1\o 2}; 1; {x_2-x_1 \o x_2-x_3}\right),
\eea
(see \cite{Matsuda} for related representations of ${da_D / du}$, 
${da / du}$, $a_D$ and $a$ in terms of the hypergeometric functions),   
the behavior of the effective theta parameter given in Eq.(\ref{thetasem}) 
follows from the known asymptotic behavior of the hypergeometric 
functions. 

These asymptotics can however be obtained directly by evaluating the 
contour integrations in Eq.(\ref{contdaddabis}) in the semiclassical
limit; $|u|, |m_i^2| \gg \Lambda^2$. In the semiclassical region the 
branch points $x_1, x_2, x_3$  are approximately given by: 
\bea
& &x_{1,2} \sim { m \pm (m^2 -u)^{1/2} \o 4u} \, \Lambda_1^3
\quad (N_f=1), \non \\
& &x_{1,2} \sim 
{ m_1 m_2 \pm (m_1^2-u)^{1/2}(m_2^2-u)^{1/2} \o 8 u}\Lambda_2^2
\quad (N_f=2), \non \\
& &x_{1,2} \sim 
{ m_1 m_2 m_3 \pm  (m_1^2-u)^{1/2}(m_2^2-u)^{1/2}(m_3^2-u)^{1/2} \o 8 u}
\Lambda_3
\quad (N_f=3),
\label{branchpt}
\eea 
and $x_3 \sim u$ for any $N_f$. The integration contours can then be  taken
so that the $\alpha$ cycle encircles $x_1$ and $x_3$, and the $\beta$ cycle
encircles $x_1$ and $x_2$, both in the counter-clockwise direction. 

Note that once these choices are made, by smoothly varying $m_i$ and/or $u$
(possibly avoiding the points where all three branch points meet and  at 
which  the low energy theory becomes superconformal invariant 
\cite{Superconf}), ${da_D / du}$ and ${da / du}$  are defined uniquely for 
any values of $u$ and $m_i$.  More precisely, they are
uniquely defined  modulo possible $SL(2,Z)$ redefinitions  needed, e.g., 
when a "quark singularity" (where a $(0,1)$ particle is massless) at large 
$m_i$ turns into a dyon singularity (where a $ (n_m, n_e)$ particle is 
massless) at small $m_i$, or when one loops around  some singularities 
(monodromy transformations).   

In order to actually evaluate the integrals it is convenient to change the 
variable from $x$ to $ \xi= (x-x_2)/(x_1 - x_2)$, and introduce 
$\xi_u = (x_3 -x_2)/(x_1 - x_2) \sim u /(x_1 - x_2)$.
Then the integral for $da/du $ can be immediately evaluated in 
the semiclassical limit as
\beq 
{da \o du} \sim  2{\sqrt{2} \o 8 \pi}{1 \o \sqrt{u}}
\int_0^1 d\xi\, {1 \o \sqrt{\xi (1 - \xi)}} = 
  {1 \o 2}{1 \o \sqrt{2u}}.
\eeq
One finds then
\beq   
a(u) = {1 \o 2} \sqrt{ 2 u} \, ( 1 + O({1 \o u}) ), 
\eeq
where an appropriate integration constant, as in Eq.(\ref{adea}), has 
to be taken into account.

 $da_D/du$ is also easily evaluated as
\beq 
{da_D \o du} \sim
2i{\sqrt{2} \o 8\pi}{1 \o (x_1 - x_2)^{1/2}}
\int_1^{\xi_u}  d\xi \,{ 1 \o  \sqrt{\xi(\xi-1)(\xi_u - \xi)}}
\sim i{\sqrt{2} \o 4\pi}{1 \o \sqrt{u}} \{\log \xi_u + 4\log 2 \}.
\eeq
After substituting the explicit values of $x_i$ given in 
Eq.(\ref{branchpt}) for $\xi_u$
the asymptotic behavior  of $\tau$ for $N_f=1, 2, 3$ is found to be:
\beq 
\tau = {da_D \o da } \simeq 
{ i \o \pi } \left\{ 
4\log a - { 1\o 2}\sum_{i=1}^{N_f}\log(a^2 - {m_i^2 \o 2}) 
-{ N_f \pi \o 2} i  + (8 - {N_f \o 2})\log 2  
\right\} + O\left({1 \o a }\right).
\eeq

The contour-integral representation for $a_D(u)$ and $a(u)$ involve
appropriate integration constants. For $N_f=1$  integration of 
Eq.(\ref{contdaddabis}) yields, apart from an  integration constant, 
\[
I_{\alpha, \beta}={\sqrt{2} \o 8\pi} \oint_{\alpha, \beta}
dx\,{-2y \o x^2},
\]
with the same contours as used for $da_D/du$ and $da/du. $
The contours are chosen so as  to avoid  the pole at $x=0$.
Note that this made no difference for  Eq.(\ref{contdaddabis}) but here such a
specification is important. 
If we evaluate $I_{\beta}$ in the asymptotic region as in the case of 
$da/du$,  we find  that $I_{\beta} \sim \sqrt{2u}/2 - m/\sqrt{2}$, 
which does not quite satisfy the boundary condition, 
$a(u)\simeq {(1 / 2)} \sqrt{ 2 u}$.
Adding therefore an appropriate integration constant, one gets,  for $N_f=1$:
  \beq 
a(u)=-{ \sqrt{2} \o 4 \pi} 
\oint_{\beta}{dx \,y \o x^2 }  +   {m \o  \sqrt{2}}.  
\label{integconsta}\eeq
(Alternatively the $\beta$ cycle can be taken so as to encircle the
pole at $x=0$;  in that case  the constant term ${m / \sqrt{2}}$ 
should be dropped.)
Note that the formula Eq.(\ref{integconsta}) has a simple  interpretation.
At the  semiclassical singularty where the two smaller branch points meet 
the $\beta$ cycle   defining $a$ shrinks to a point.  The additive constant in 
Eq.(\ref{integconsta}) is there just to ensure that such a singularity 
occurs at $a= m / \sqrt{2}.$

Similarly, the integration constant in the cases of $N_f=2$ and $N_f=3$ can be
determined from the requirement that the semiclassical singularity
($u_i$) corresponding to  the $i$-th massless quark (at which the $\beta$ 
cycle is defined to shrink to zero)  occurs at  $a= m_i / \sqrt{2}.$  
Taking the first quark as the reference system  one finds, for all $N_f$, 
that
\beq 
a(u)= \oint_{\beta} \lambda_{SW} +   {m_1 \o  \sqrt{2}},  
\eeq
where 
\beq   
\lambda_{SW}= -{\sqrt{2} \o 4 \pi} 
{dx \, y \o x^2 - \Lambda_2^4/64}   
\eeq
for $N_f=2$; 
$\lambda_{SW}$ for $N_f=3$ has been recently given in a simple form by 
{\'A}lvarez-Gaum\'e, Marino and Zamora\cite{AMZ}:
\beq   
\lambda_{SW}= -{\sqrt{2} \o 16  \pi} 
{dx \o y }(2 x - 4 u - \sum_{n=1}^4 { y_n x_n \o x -x_n}),  
\eeq
where
\bea 
&&  y_1= u-m_1 m_2 -m_1 m_3 +  m_3 m_2 -x_1; \quad 
x_1=c ( -m_1 +m_2 +m_3), \non \\ 
&& y_4= u+ m_1 m_2 + m_1 m_3 +  m_3 m_2 -x_4; \quad 
x_4=c ( -m_1-m_2 - m_3), 
\eea 
and $y_2=y_1(m_1 \leftrightarrow m_2)$, 
$y_3=y_1(m_1 \leftrightarrow m_3)$, and $c\equiv \Lambda_3/8$.  
(Note some minor misprints  in Eq.(2.60) and Eq.(2.62) of \cite{AMZ}.) 

As one moves from the first singularity $u_1$  to the second quark 
singularity, $u_2$, the integration over the $\beta$ cycle picks up the 
residue at a pole, such that  $a(u_2)= {m_2 / \sqrt{2}},$ and so on. 

Finally, the integration constant for $a$ has been discussed in 
Sec.\ref{sec:qnum} One has 
\beq 
a_D(u)=-{ \sqrt{2} \o 4 \pi} 
\oint_{\alpha} {dx \, y \o x^2  }  + \sum_{i=1}^{N_f} {m_i \o 2 \sqrt{2}}, 
\eeq
where the $\alpha$ cycle is such that at one of the $(1,0)$ singularities
it shrinks to zero. 


\end{document}